\documentclass[final,5p,times,twocolumn]{elsarticle}

\usepackage{amssymb}
\usepackage{amsmath}
\usepackage{lineno}
\usepackage{listings}
\usepackage{dcolumn}
\usepackage{graphicx}
\usepackage{array}
\usepackage{color}

\usepackage{hyperref}
\hypersetup{colorlinks, linkcolor=blue}

\newcounter{bla}

\makeatletter
\def\ps@pprintTitle{%
  \let\@oddhead\@empty
  \let\@evenhead\@empty
  \let\@oddfoot\@empty
  \let\@evenfoot\@oddfoot
}
\makeatother

\begin{document}

% -- BEGIN: CUSTOMIZE LISTINGS ENVIRONMENT
\definecolor{codegreen}{rgb}{0,0.6,0}
\definecolor{codegray}{rgb}{0.5,0.5,0.5}
\definecolor{codepurple}{rgb}{0.58,0,0.82}
\definecolor{backcolour}{rgb}{0.95,0.95,0.92}
 
\lstdefinestyle{mystyle}{
    %backgroundcolor=\color{backcolour},   
    commentstyle=\color{codegreen},
    keywordstyle=\color{magenta},
    numberstyle=\tiny\color{codegray},
    stringstyle=\color{codepurple},
    basicstyle=\ttfamily\footnotesize,
    breakatwhitespace=false,         
    breaklines=true,                 
    captionpos=b,                    
    keepspaces=true,                 
    numbers=left,                    
    numbersep=5pt,                  
    showspaces=false,                
    showstringspaces=false,
    showtabs=false,                  
    tabsize=2
}
 
\lstset{style=mystyle}
% -- END: CUSTOMIZE LISTINGS ENVIRONMENT

\sloppy
\begin{frontmatter}

\title{SWtools: A Python module implementing iterative solvers for soliton solutions of nonlinear Schrödinger-type equations}

\author[add1,add2]{O. Melchert\corref{mycorrespondingauthor}}
\author[add1,add2]{A. Demircan}

\cortext[mycorrespondingauthor] {Corresponding author.\\\textit{E-mail address:} melchert@iqo.uni-hannover.de}
\address[add1]{Leibniz Universit\"at Hannover, Institute of Quantum Optics, Welfengarten 1, 30167 Hannover, Germany}
\address[add2]{Leibniz Universit\"at Hannover, Cluster of Excellence PhoenixD, Welfengarten 1A, 30167 Hannover, Germany}

\begin{abstract}
Solitons are ubiquitous in nature and play a pivotal role in the structure and dynamics of solutions of nonlinear propagation equations. 
In many instances where solitons exist, analytical expressions of these special objects are not available.
The presented software fills this gap, allowing users to numerically calculate soliton solutions for nonlinear Schrödinger-type equations by iteratively solving an associated nonlinear eigenvalue problem.
The package implements a range of methods, including the spectral renormalization method (SRM), and a relaxation method for the problem with additional normalization constraint (NSOM). 
We verify the implemented methods in terms of a problem for which an analytical soliton expression is available, and demonstrate the implemented functionality by numerical experiments for example problems in nonlinear optics and matter-wave solitons in quantum mechanics.
The presented Python package is open-source and released under the MIT License in a publicly available software repository (\url{https://github.com/omelchert/SWtools}).


\end{abstract}
\end{frontmatter}

\tableofcontents

\section{Introduction\label{sec:intro}}
%
%\paragraph{Introduction}
%
% -- TOPICAL INTRODUCTION
% ... SOLITARY WAVES
\emph{Solitary waves} (SWs) describe localized solutions of nonlinear wave equations \cite{Zabusky:PRL:1965,Zakharov:JETP:1972,Drazin:BOOK:1989}.
They travel with constant shape and speed and can only exist when the linear effects on the pulse envelope are balanced by the effects imposed by the nonlinearity. 
% ... COLLISIONS
Collisions among SWs are typically inelastic \cite{Kivshar:RMP:1989,Frauenkron:PRE:1996,Jakubowski:PRE:1997,Anastassiou:PRL:1999,Yang:PRL:2000,Dmitriev:Chaos:2002,Feigenbaum:OE:2004,Goodman:PRL:2007,Edmonds:NJP:2017,Dingwall:NJP:2018,Melchert:PRL:2019,Rao:PRE:2020,Melchert:PRA:2024}: the colliding pulses can exchange energy and momentum, may suffer radiation losses, and even form short-lived bound states. 
%
% ... SOLITONS
However, special SWs -- called \emph{solitons} -- exist \cite{Zabusky:PRL:1965}, that interact elastically and emerge from collisions with unchanged shape and speed.
The mutual interaction of such true solitons is evidenced only by a phase-shift that both pulses acquire during their collision~\cite{Zakharov:JETP:1972}.
%
% TODO: weitere Beispiele wieder einfügen
%
SWs and solitons arise in diverse fields of physics, governed, e.g., by 
the Korteweg-deVries equation in hydrodynamics and plasma physics \cite{Zabusky:PRL:1965},
%the Born-Infeld equation in nonlinear electrodynamics \cite{},
% Burgers equation in applied mathematics \cite{},
the Schrödinger-Poisson system in computational cosmology \cite{Paredes:PD:2020,Zagorac:PRD:2022},
the Gross-Pitaevskii equation in quantum mechanics \cite{Dalfovo:RMP:1999,Edmonds:NJP:2017},
and nonlinear Schrödinger-type equations in nonlinear optics \cite{Mitschke:BOOK:2016}.
They play an important role in the structure and dynamics of general solutions of the underlying equations. %
%
% \cite{Kivshar:RMP:1989,Dudley:RMP:2006}
%\cite{Kivshar:RMP:1989,Frauenkron:PRE:1996,Jakubowski:PRE:1997,Anastassiou:PRL:1999,Yang:PRL:2000,Dmitriev:Chaos:2002,Feigenbaum:OE:2004,Goodman:PRL:2007,Edmonds:NJP:2017,Dingwall:NJP:2018,Rao:PRE:2020,Melchert:PRA:2024}.
%
%Here, our interest lies in the latter field, where \emph{optical} solitons and soliton related phenomena do not cease to offer new and exciting perspectives~\cite{Redondo:NP:2023}.
%
%Originally considered as information carriers in fiber optical communication systems \cite{Hasegawa:APL:1973,Hasegawa:APL:1973_2,Mollenauer:PRL:1980}, they may now even guide the way for novel applications in ultrafast optical science \cite{Corkum:NP:2007,Krausz:NP:2014} which is currently advancing towards the domain of attosecond physics \cite{Krausz:RMP:2009}.
%
Despite the difference between SWs and solitons, we will subsequently use both terms interchangeably.

%
% \item \textit{Explain why the software is important and describe the exact (scientific) problem(s) it solves.}
%
% TODO: 2 Sätze
%
Soliton propagation is of considerable scientific and technological interest \cite{Mitschke:BOOK:2016,Redondo:NP:2023}. The task of finding soliton solutions for nonlinear wave equations, however, is a nontrivial task.
Mathematical techniques have been developed that yield analytical soliton solutions for several paradigmatic nonlinear equations \cite{Zakharov:JETP:1972,Gardner:PRL:1967,Lax:CPAM:1968,Ablowitz:PRL:1973,Ablowitz:SAM:1974,Satsuma:PTP:1974,Kaup:JMP:1975,Miles:SIAM:1981}. 
For most nonlinear problems, however, one has to rely on numerical techniques to determine the shapes of solitons. 
%
% \item \textit{Introduce related work in literature (cite or list algorithms used, other software etc.).}
%
% Let us point out that other methods for the numerical calculation of SW solutions in nonlinear systems exist, such as, e.g.,
%%
%For the numerical solution of linear problems iterative methods have been used extensively \cite{Press:BOOK:2007,Langtangen:BOOK:2019}.
%
%Analogous methods have been generalized to nonlinear problems as well \cite{Schechter:AMS:1962,Ortega:SIAM:1966,Brewster:NM:1984,Yang:BOOK:2010,Langtangen:BOOK:2019}. 
%
For this purpose, iterative methods, originally developed for linear problems \cite{Press:BOOK:2007,Langtangen:BOOK:2019}, have been generalized to nonlinear settings \cite{Schechter:AMS:1962,Ortega:SIAM:1966,Brewster:NM:1984,Yang:BOOK:2010,Langtangen:BOOK:2019}. 
Commonly used methods are, e.g.\ the shooting method \cite{Haelterman:PRE:1994,Mitchell:PRL:1997,Yang:PRE:2002}, Petviashvili-type iteration methods \cite{Petviashvili:SJPP:1976,Lakoba:JCP:2007}, imaginary time evolution methods \cite{Bao:SIAM:2004,Lehtovaara:JCP:2007,Yang:SIAM:2008}, squared-operator methods \cite{Yang:SIAM:2007}, Newton's method \cite{Boyd:JCP:2002,Maytee:PRA:2006,Melchert:SFX:2021}, and conjugate gradient methods~\cite{Yang:JCP:2009,Lakoba:PD:2009}. 
We here provide a Python framework that allows to conveniently implement such iterative methods for the numerical computation of SW solutions for a nonlinear Schrödinger-type equation.

% -- Final paragraph of introductory part
The article is organized as follows.
In Section~\ref{sec:problem} we state the computational problem solved by {\tt{SWtools}}.
In Section~\ref{sec:software} we detail the implemented methods.
In Section~\ref{sec:examples} we demonstrate different aspects of the software: we report a verification test and show the effectiveness of the implemented algorithms for various example problems in nonlinear optics and matter-wave solitons in quantum mechanics.
We comment on impact in Section~\ref{sec:impact}, and conclude in Section~\ref{sec:conclusion}.

%
%Considering parameter ranges beyond those of the known meta-envelope solitons, we compute these complex wavelet-like solutions numerically by the spectral renormalization method.
%
%Especially in nonlinear optics, where they have been considered as information carriers in fiber optical communication systems \cite{Hasegawa:APL:1973,Hasegawa:APL:1973_2,Mollenauer:PRL:1980}, \emph{optical} solitons and soliton related phenomena do not cease to offer new and exciting perspectives~\cite{Redondo:NP:2023}.
%
%where the NSE models the combined effects of group-velocity dispersion (GVD) and third-order nonlinearity, 
%
%Here we consider a higher-order NSE (HONSE) for the pulse envelope $\psi\equiv A(z,\tau)$ in fiber-optic notation, including second-order dispersion (2OD), third-order dispersion (3OD), and fourth-order dispersion (4OD) in the form
%\begin{align}
%i \partial_z A = \frac{\beta_2}{2} \partial_\tau^2 A + i\frac{\beta_3}{6}\partial_\tau^3 A - \frac{\beta_4}{24} \partial_\tau^4 A - \gamma |A|^2 A, \label{eq:HONSE}
%\end{align}
%where $z$ is the propagation coordinate, $\tau=t-\beta_1 z$ is a retarded time, $\beta_1$ is the inverse group-velocity at the carrier frequency, and $\gamma$ is the nonlinear parameter. 
%

% -- EXACT SCIENTIFIC PROBLEM SOLVED BY THE SOFTWARE
%
\section{Computational problem solved by the software \label{sec:problem}}
We consider the generalized nonlinear Schrödinger equation (GNSE) for a complex envelope $\psi\equiv \psi(\eta,\xi)$, given by
\begin{align}
i \partial_\eta\,\psi = -\hat{L}(i\partial_\xi)\,\psi - F[|\psi|^2, \xi]\,\psi, \label{eq:GNLS}
\end{align}
where $\eta$ is the propagation distance, $\xi$ is a one-dimensional transverse coordinate, $\hat{L}$ is a linear differential operator, and $F$ is a real-valued nonlinear functional. 
We seek SW solutions of Eq.~(\ref{eq:GNLS}) using the stationary-state Ansatz
\begin{align}
\psi(\eta,\xi)=U(\xi)\,e^{i\kappa \eta}, \label{eq:ansatz}
\end{align}
where the possibly complex function $U$ satisfies the nonlinear eigenvalue problem (NEVP)
\begin{align}
\kappa\,U = \hat{L}(i\partial_\xi)\,U + F[|U|^2, \xi]\,U, \label{eq:EVP}
\end{align}
subject to the boundary conditions $U\to 0$ as \mbox{$|\xi|\to \infty$}.
Solutions~(\ref{eq:ansatz}) of Eq.~(\ref{eq:EVP}) are called nonlinear bound states, SWs, or stationary states. The parameter $\kappa$ specifies the eigenvalue, or propagation constant of the  solution.
Evaluated for a solution $\psi$, the functionals
\begin{subequations}\label{eq:IOM}
\begin{align}
N[\psi] &= \int |\psi|^2~{\rm{d}}\xi, \label{eq:IOM_N}\\
H[\psi] &= \int (\psi^*\,\hat{L}\,\psi + F[|\psi|^2, \xi]\,|\psi|^2 )~{\rm{d}}\xi, \label{eq:IOM_H} 
\end{align}
\end{subequations}
are assumed to be finite and independent of the propagation distance $\eta$, and satisfy  $H[U]=\kappa\, N[U]$.
Let us emphasize that in order to ensure existences, uniqueness, and stability of solutions~(\ref{eq:ansatz}), the right-hand-side of Eq.~(\ref{eq:GNLS}) can be subject to further case-related conditions \cite{Rose:PD:1988,Zakharov:JETP:1998,Yang:SIAM:2008,Tsoy:PRA:2024}.
For specificity, we assume a linear differential operator
\begin{align}
\hat{L}(i\partial_\xi)= c_1 (i\partial_\xi) + c_2 (i \partial_\xi)^2 + c_3 (i\partial_\xi)^3 + c_4 (i \partial_\xi)^4, \label{eq:L}
\end{align}
with real-valued coefficients $c_L=(c_1,\ldots,c_4)$,
allowing to address key models in quantum mechanics \cite{Besse:BOOK:2015} and nonlinear photonics \cite{Agrawal:BOOK:2019}.

The provided software allows a user to numerically obtain nonlinear bound states~(\ref{eq:ansatz}) of the GNSE~(\ref{eq:GNLS}).
Therefore, {\tt{SWtools}} implements two complementary methods that iteratively solve the NEVP~(\ref{eq:EVP}), starting from user-supplied initial conditions.
%
% -- Spectral renormalization method
%
The first method, detailed in Section~\ref{sec:alg_02}, implements a spectral renormalization technique~\cite{Ablowitz:PD:2003,Ablowitz:OL:2005,Musslimani:JOSAB:2004,Ablowitz:EPJST:2009}. It solves Eq.~(\ref{eq:EVP}) for the function $U$ for a given eigenvalue $\kappa$.
This method works mainly in the Fourier domain associated with the coordinate $\xi$, exhibits especially high accuracy, and is easy to customize to common problems in nonlinear optics that can be cast in the form of the GNSE~(\ref{eq:GNLS}).
%
% -- Successive overrelaxation method (SOR)
%
The second method, detailed in Section~\ref{sec:alg_01}, implements a relaxation technique \cite{Brewster:NM:1984,Press:BOOK:2007}. It solves Eq.~(\ref{eq:EVP}) for both, the function $U$ and eigenvalue $\kappa$, in presence of an additional normalization constraint $N[U]=N_0$.
Such a problem is posed, e.g., by the Gross-Pitaevskii equation (GPE) for ground-state solutions of Bose-Einstein condensates (BECs) \cite{Bao:SIAM:2004,Bao:JCP:2003,Liu:SIAM:2021}.
As additional feature, {\tt{SWtools}} allows to obtain the ground-state of the nonlinear problem and excited states by propagating different trial functions in successive manner and using an orthogonalization method to eliminate any overlap with previously found solutions. 
For a concise presentation, we consider a one-dimensional ($d=1$) transverse coordinate $\xi$.
The methods provided with {\tt{SWtools}} can, however, be extended to $d$-dimensional transverse coordinates $\vec{\xi}=(\xi_1, \ldots, \xi_d)$ and to more complex instances of the linear operator $\hat{L}$ in straight forward manner. 

%
%The second method (Sect.~\ref{sec:alg_02}) implements a spectral renormalization technique~\cite{Ablowitz:OL:2005,Ablowitz:EPJST:2009}. 
%
%It works mainly in the Fourier domain associated with the coordinate $\xi$, exhibits especially high accuracy, and is easy to customize to common problems in nonlinear optics that can be described by some variant of the GNSE~(\ref{eq:GNLS}).
%
%It employs the Fourier representation of $\hat{L}$, providing especially high accuracy effectively independent of the meshwidth~\cite{Trefethen:BOOK:2000}.
%
% It works in the Fourier domain associated to the coordinate $\xi$,
% and solves derivatives in $\hat{L}$ via Fourier spectral differentiation,
%
% \item \textit{Indicate in what way the software has contributed (or will contribute in the future) to the process of scientific discovery; if available, please cite a research paper using the software.}
%

%
% \item \textit{Provide a description of the experimental setting. (How does the user use the software?)}
%
% TODO

%\clearpage

\section{Software description\label{sec:software}}
%
%\textit{Describe the software. Provide enough detail to help the reader understand its impact. }
%
%
{\tt{SWtools}} is openly available \cite{SWtools:GH:2025}, hosted on the code development platform GitHub (\url{https://github.com}).
It provides an extendible software framework for iterative methods that enable a user to calculate solutions to the NEVP~(\ref{eq:EVP}), and allows to treat two variants of the problem: 
%
%The Python toolbox {\tt{SWtools.py}} implements iterative methods that enable a user to calculate nonlinear bound states of the GNSE~(\ref{eq:GNLS}) that satisfy Ansatz~(\ref{eq:ansatz}).
%
%Formally, it allows to treat two variants of the corresponding NEVP~(\ref{eq:EVP}): 
%
\begin{enumerate}
\item The ``bare'' NEVP, where a solution $U$ for a given eigenvalue $\kappa$ is computed;
\item The ``constraint'' NEVP, with Eq.~(\ref{eq:EVP}) supplemented by a normalization constraint $N[U]=N_0$ [see Eq.~(\ref{eq:IOM_N})], where a solution $U$ for a given norm
$N_0$ is computed. In this case, however, the eigenvalue $\kappa$ is \emph{a priori} unknown and needs to be estimated on the go.
\end{enumerate}
While the bare NEVP is frequently encountered in modeling and computation in nonlinear optics \cite{Ablowitz:OL:2005,Ablowitz:EPJST:2009}, the constraint NEVP is encountered in the context of BECs \cite{Besse:BOOK:2015}. 
{\tt{SWtools}} implements tested and reliable algorithms for both variants of the NEVP.

\begin{figure}[!t]
\includegraphics[width=\linewidth]{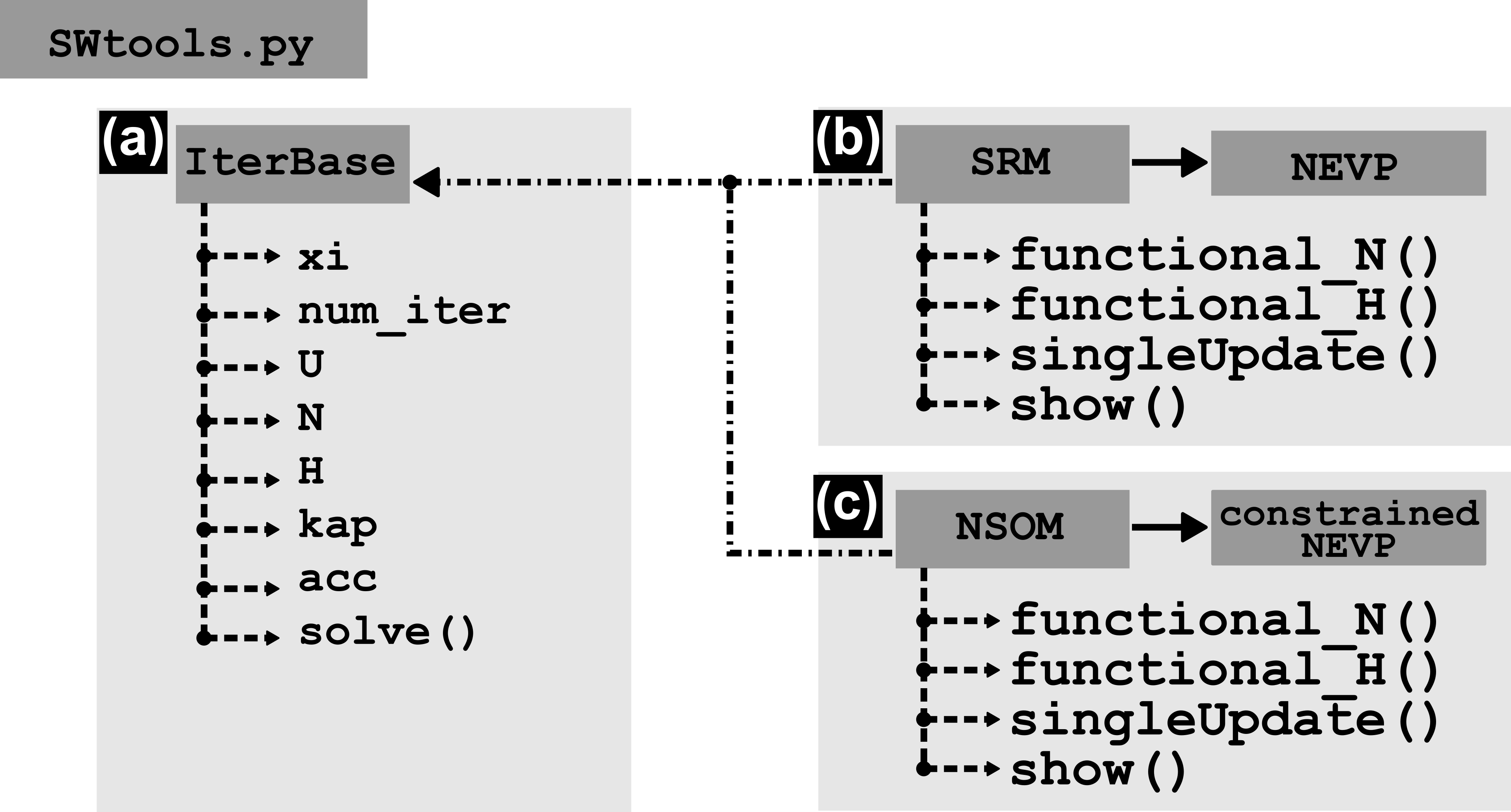}
\caption{
Pictorial outline of {\tt{SWtools}}.
(a) Superclass {\tt{IterBase}}.
(b) Subclass {\tt SRM} implementing the spectral renormalization method for solving Eq.~(\ref{eq:EVP}), see Section~\ref{sec:alg_02} and \ref{sec:SRM_subclass}.
(c) Subclass {\tt NSOM} implementing the nonlinear successive overrelaxation method for solving the NEVP~(\ref{eq:EVP}) with normalization constraint, see Section~\ref{sec:alg_01} and \ref{sec:NSOM_subclass}.
Solid arrow indicates class instantiation (``creates'' relationship),
dashed arrows indicate data structures (``has-a'' relationship),
dash-dotted arrow indicates reference to an object (``is-a'' relationship).
\label{fig:pictorialOverview}}
\end{figure}

\subsection{Software architecture}
%
%\textit{  Give a short overview of the overall software architecture; provide a pictorial overview where possible; for example, an image showing the components. If necessary, provide implementation details.}
%
%\paragraph{Overview}
%
{\tt{SWtools}} is written using the Python programming language \cite{Rossum:1995}, and depends on the functionality of numpy, scipy \cite{Virtanen:NM:2020}, and matplotlib \cite{Hunter:CSE:2007}. 
It can be cloned directly from it's GitHub repository, where it is available under a MIT license \cite{SWtools:GH:2025}. 
A pictorial outline of the {\tt{SWtools}} Python package is shown in Fig.~\ref{fig:pictorialOverview}.
In the current version it features:
\begin{itemize}
\setlength{\itemsep}{0em}
\item A class data structure {\tt{IterBase}} [Fig.~\ref{fig:pictorialOverview}(a)], constituting a superclass for the implemented methods. It provides the class method {\tt{solve()}}, which organizes the iterative solution of the NEVP~(\ref{eq:EVP}), and keeps track of intermediate solutions;
\item The subclass {\tt{SRM}} [Fig.~\ref{fig:pictorialOverview}(b), Section~\ref{sec:alg_02}], which implements the spectral renormalization method (SRM) of Refs.~\cite{Ablowitz:PD:2003,Ablowitz:OL:2005,Musslimani:JOSAB:2004,Ablowitz:EPJST:2009} for the bare NEVP;  
\item The subclass {\tt{NSOM}} [Fig.~\ref{fig:pictorialOverview}(c), Section~\ref{sec:alg_01}], which implements a nonlinear successive overrelaxation method (NSOM) for the constraint NEVP.
\end{itemize}
%
% -- implementation details IterBase
%
%\paragraph{Implementation details}
In our design of the data structures in {\tt{SWtools}}, we decided to bundle all solution-characteristic data, i.e.\ data a user might want to routinely interact with, at the class {\tt{IterBase}}. 
This comprises the set of data shown in Fig.~\ref{fig:pictorialOverview}(a), subsequently listed in the format ``data (class attribute): description'': 
\begin{itemize}
\setlength{\itemsep}{0em}
\item $\xi$ ({\tt{xi}}): Discrete transverse coordinate given by a uniform grid $\xi=(\xi_0, \ldots, \xi_M)$ with mesh size $\Delta \xi$;
\item $n^\star$ ({\tt{num\_iter}}): Performed number of iterations;
%\item $U^\star$ ({\tt{U}}): Solution after iteration step $n^\star$;
\item $U^\star$ ({\tt{U}}): Solution  $U^{\star}=(U_0^{\star}, \ldots, U_M^{\star})$ after step $n^\star$;
\item $N^\star$ ({\tt{N}}): Value of functional~(\ref{eq:IOM_N}) after step $n^\star$;
\item $H^\star$ ({\tt{H}}): Value of functional~(\ref{eq:IOM_H}) after step $n^\star$;
\item $\kappa$ ({\tt{kap}}): Eigenvalue in Eq.~(\ref{eq:EVP});
\item $\epsilon^\star$ ({\tt{acc}}): Terminal accuracy $\epsilon_{n^\star}$.
\end{itemize}
%
%Given two solutions $U_{n-1}$ and $U_n$ at successive iteration steps ($n>0$), the accuracy $\epsilon_n$ at step $n$ is calculated as 
%\begin{align}
%\epsilon_{n} = \frac{\lVert U_n-U_{n-1} \rVert}{\lVert U_{n-1}\rVert}, \label{eq:acc}
%\end{align}
%with norm $\lVert a \rVert = (\int |a|^2~{\rm{d}}\xi)^{1/2}$.
%
Given two solutions $U^{(n-1)}$ and $U^{(n)}$ at successive iteration steps ($n>0$), the accuracy $\epsilon_n$ at step $n$ is calculated as
\begin{align}
\epsilon_{n} = \lVert U^{(n-1)}-U^{(n)} \rVert, \label{eq:acc}
\end{align}
with discrete $\ell^2$-norm $\lVert a \rVert = \left(\sum_m |a_m|^2~\Delta\xi\right)^{1/2}$.
%
%the $\ell_2$-norm of the difference between both functions.
%
%Let us point out that in case of the constrained NEVP, the value of $\kappa$ that can be accessed through an instance of the subclass {\tt{NSOM}} is the eigenvalue estimate ${\rm{K}}^\star=H^\star/N^\star$.
%
Let us point out that in case of the constrained NEVP, the \emph{a priori} unknown value of $\kappa$ is estimated as ${\rm{K}}^\star=H^\star/N^\star$.
%
% -- implementation details subclasses
%
The logic implementing the linear operator $\hat{L}$, and the nonlinear functional $F$ is bundled at the subclass level, see Figs.~\ref{fig:pictorialOverview}(b,c).
We consider this a reasonable design decision since the precise implementation can be method specific. For instance, while the subclass {\tt{SRM}} implements a Fourier-representation of the operator $\hat{L}$, subclass {\tt{NSOM}} implements all derivatives in $\hat{L}$ by their corresponding five-point finite-difference approximation.
A reference manual, containing a description of the base class {\tt{IterBase}} as well as an example showing how to extend the functionality of {\tt{SWtools}} by subclassing is provided online \cite{SWtools:GH:2025}.
 
\subsection{Software functionalities}
%
%\textit{  Present the major functionalities of the software.}

{\tt{SWtools}} provides a spectral renormalization method for the bare NEVP, detailed in Section~\ref{sec:alg_02}, and a relaxation method for the constraint NEVP, see Section~\ref{sec:alg_01}.
It allows users to specify custom measures of accuracy different from Eq.~(\ref{eq:acc}), and visualize a solution along with its convergence information.
%
%implements the functionals~(\ref{eq:IOM}) for assessing the properties of the found solutions,
%
%Additional methods for the solution of Eq.~(\ref{eq:EVP}) can be implemented by simply deriving classes from {\tt{IterBase}}, allowing developers to reuse our tested software components.
%
Let us note that in our computational research projects, which are usually carried out by scripting, we import the functionality of {\tt{SWtools}} into more specific project code.
The short workflows presented in Section~\ref{sec:examples} reflect this.
To facilitate a simple presentation, we consider custom applications in quantum mechanics and nonlinear photonics where the transverse coordinate $\xi$ is one-dimensional ($d=1$).
%
% The methods provided with {\tt{SWtools}} can, however, be extended to $d$-dimensional transverse coordinates $\vec{\xi}=(\xi_1, \ldots, \xi_d)$ and to more complex instances of the linear operator $\hat{L}$ in straight forward manner. 
%
Extension of the functionality of {\tt{SWtools}} to treat the NEVP for a problem with two-dimensional ($d=2$) transverse coordinate $\vec{\xi}=(\xi_1,\xi_2)$ is demonstrated by the additional Python module {\tt{SWtools\_ext\_SRM2D.py}} (provided along with {\tt{SWtools}} under Ref.~\cite{SWtools:GH:2025}), which implements a $d=2$ SRM \cite{Musslimani:JOSAB:2004}.
%
%This is demonstrated in terms of an additional Python module {\tt{SWtools\_ext\_SRM2D.py}}, which extends the functionality of the {\tt{SWtools}} base package by implementing a two-dimensional ($d=2$) SRM for $\vec{\xi}=(\xi_1, \xi_2)$, see the example in Section~\ref{sec:SRM2D}.
%
%The software can be used on its own, or as extension module for {\tt{GNLStools}} \cite{Melchert:SFX:2022} and {\tt{py-fmas}} \cite{Melchert:CPC:2022}, allowing to study the propagation dynamics of the obtained solutions in terms of the initial-value problem for Eq.~(\ref{eq:GNLS}).
%
Below we describe the implemented methods and cover the main features of the corresponding data structures. Numerical experiments that verify the implementation are reported in Section~\ref{sec:verification}.

%\clearpage

\subsubsection{Description of the spectral renormalization method (SRM) \label{sec:alg_02}}

Subsequently we briefly describe the spectral renormalization method thoroughly detailed in Refs.~\cite{Ablowitz:PD:2003,Ablowitz:OL:2005,Musslimani:JOSAB:2004,Ablowitz:EPJST:2009,Fibich:PD:2006}.
We therefore consider the NEVP~(\ref{eq:EVP}) in the form
\begin{subequations}
\label{eq:EVP_SRM}
\begin{align}
\kappa\,U &= \hat{L}(i\partial_\xi)\,U + F[|U|^2, \xi]\,U, \quad &&|\xi|\leq \infty, \label{eq:SRM_1}\\
%\partial_t\,A &= \left( \hat{L} + F-{\rm{K}}\right)\,A, \quad &&|\xi|\leq L,~t\geq 0,\label{eq:EVP_aux_diffeq}\\
%U(\xi) &= U_0(\xi), \quad &&|\xi| < L,\label{eq:SRM_ic}\\
U(\pm \infty) &= 0,\label{eq:SRM_bc}
\end{align}
\end{subequations}
with boundary conditions~(\ref{eq:SRM_bc}) consistent with the localization condition $U\to 0$ for $|\xi|\to \infty$.
The function $U(\xi)$ can be related to its Fourier transform $u(k)$ by means of the transform equations 
\begin{subequations} \label{eq:FT}
\begin{align}
u(k) &= {\mathcal{F}}\left[U(\xi)\right] =\int_{-\infty}^{\infty} U(\xi)\,e^{i k \xi}~{\rm{d}}\xi, \label{eq:FT_FT}\\
U(\xi) &= {\mathcal{F}}^{-1}\left[ u(k) \right] = \frac{1}{2\pi} \int_{-\infty}^{\infty} u(k)\,e^{-ik\xi}~{\rm{d}}k, \label{eq:FT_IFT}
\end{align}
\end{subequations}
implying the identity $[(i \partial_\xi)^n-k^n]\,e^{-ik \xi} = 0$.
The Fourier representation of the linear operator~(\ref{eq:L}) then reads $L(k)= c_1 k + c_2 k^2 + c_3 k^3 + c_4 k^4$.
Taking the Fourier transform of Eq.~(\ref{eq:SRM_1}) and rearranging yields a fixed-point equation for $U$, given by
\begin{align}
U = {\mathcal{F}}^{-1}\left[\frac{1}{\kappa - L(k)} {\mathcal{F}}\left[F[|U|^2,\xi]\,U\right]\right], \label{eq:SRM_FP}
\end{align}
for which a convergent iteration scheme can be derived \cite{Petviashvili:SJPP:1976,Ablowitz:EPJST:2009,Fibich:PD:2006}.
The resulting solution $U^\star$ satisfies Eqs.~(\ref{eq:EVP_SRM}), and exhibits $N[U^\star]>0$ [Eq.~(\ref{eq:IOM_N})].
Let us note that the existence of solutions to Eqs.~(\ref{eq:EVP_SRM}) may be subject to further case-related conditions for $\hat{L}$ and $F$ \cite{Rose:PD:1988,Zakharov:JETP:1998,Yang:SIAM:2008,Tsoy:PRA:2024}.
The corresponding class {\tt{SRM}}, derived from {\tt{SWtools}} base class {\tt{IterBase}} (Fig.~\ref{fig:pictorialOverview}), is documented online \cite{SWtools:GH:2025} and in \ref{sec:SRM_subclass}.

%Numerical experiments that verify the implementation are reported in Sect.~(\ref{sec:verification}).

%\clearpage

\subsubsection{Description of the nonlinear successive overrelaxation method (NSOM) \label{sec:alg_01}}
Below, we briefly describe our approach to solve the constraint NEVP~(\ref{eq:EVP}).
We rewrite Eq.~(\ref{eq:EVP}) as an auxiliary initial-boundary value diffusion problem for a function $A=A(t, \xi)$ \cite{Langtangen:BOOK:2019,Press:BOOK:2007}, respecting an auxiliary time coordinate $t$ in the form
\begin{subequations}
\label{eq:EVP_aux}
\begin{align}
%\partial_t\,A &= \left( \hat{L}(i\partial_\xi) + F[|A|^2, \xi]-\kappa\right)\,A, \quad |\xi|\leq L,~t\geq 0,\label{eq:EVP_aux_difeq}\\
\partial_t\,A &= \left( \hat{L} + F-{\rm{K}}\right)\,A, \quad &&|\xi|\leq \infty,~t\geq 0,\label{eq:EVP_aux_diffeq}\\
A(0, \xi) &= U_0(\xi), \quad &&|\xi| < \infty,\label{eq:EVP_aux_ic}\\
A(t,\pm \infty) &= 0, \quad &&t>0,\label{eq:EVP_aux_bc}\\
N[A] &= N_0, \quad &&t>0.\label{eq:EVP_aux_N}
\end{align}
\end{subequations}
Above, $U_0(\xi)$ is a user-supplied initial condition. $N_0$ is a user-supplied value for the normalization constraint~(\ref{eq:EVP_aux_N}), assumed to control the evolution of $A$. 
In Eq.~(\ref{eq:EVP_aux_diffeq}) we estimate the eigenvalue $\kappa$, which is \emph{a priori} unknown, as ${\mathrm{K}}=H[A]/N[A]$ [see Eqs.~(\ref{eq:IOM})].
We employ the boundary condition~(\ref{eq:EVP_aux_bc}), consistent with the localization condition $A(t,\xi)\to 0$ for $|\xi|\to \infty$. 
For the solution of Eqs.~(\ref{eq:EVP_aux}), we derived a custom nonlinear relaxation method using a Gauss-Seidel update with successive overrelaxation \cite{Langtangen:BOOK:2019,Langtangen:BOOK:2017,Press:BOOK:2007}.
In our implementation we employ an explicit time integration method for Eq.~(\ref{eq:EVP_aux_diffeq}), making it easy to handle the nonlinear functional $F$, and we substitute all derivatives in $\hat{L}$ by their corresponding five-point approximation.
The time increment is chosen dynamically at each iteration step and entirely prescribed in terms of the considered discretization. It can thus be kept entirely ``hidden'' from the user. 
Using this method, a user-specified initial condition $A(0,\xi)=U_0(\xi)$ relaxes to a stationary solution as $t\to\infty$.
In this method, successive iteration steps correspond to successive points in time.
Again, we note that the existence of solutions to Eqs.~(\ref{eq:EVP_aux}) may be subject to further case-related conditions for $\hat{L}$ and $F$ \cite{Rose:PD:1988,Zakharov:JETP:1998,Yang:SIAM:2008,Tsoy:PRA:2024}.
The corresponding class {\tt{NSOM}}, derived from {\tt{SWtools}} base class {\tt{IterBase}} (Fig.~\ref{fig:pictorialOverview}), is documented online \cite{SWtools:GH:2025} and in \ref{sec:NSOM_subclass}.

% ... CONDITIONS:
% [Yang:SIAM:2008] -- general linear self-adjoint semi-negative-definite constant-coefficient pseudo-differential operator.

%\cite{Rose:PD:1988,Zakharov:JETP:1998,Yang:SIAM:2008,Tsoy:PRA:2024}.

%\clearpage

% \subsection{Sample code snippets analysis (optional)}

\section{Usage examples\label{sec:examples}}
%
%\textit{Provide at least one illustrative example to demonstrate the major functions of your software/code.}
%
%Below we illustrate the major functions implemented by {\tt{SWtools}}.  
%
Below we show examples that demonstrate the function of {\tt{SWtools}} and help a user to become familiar with its features. 
Python scripts that allow to reproduce all subsequent examples are provided online \cite{SWtools:GH:2025}.
%
%All scripts shown as code-listings are provided along with the code under \cite{SWtools:GH:2025}.
%
In Section~\ref{sec:verification} we report a verification test based on a known analytical solution for a higher-order nonlinear Schrödinger equation (HONSE).
Sections~\ref{sec:1DBEC}-\ref{sec:nlinMicro} demonstrate the calculation of ground-states for a GPE, soliton solitons for a HONSE, and nonlinear bound states for a GNSE with inhomogeneous nonlinearity, respectively.
To demonstrate the calculation of nonlinear excited states, we consider a GPE with harmonic trapping potential in Section~\ref{sec:excitedStates}.
A workflow that shows software integration of {\tt{SWtools}} with {\tt{py-fmas}} \cite{Melchert:CPC:2022}, an open-source Python package for the accurate simulation of the propagation dynamics of optical pulses, is detailed in Section~\ref{sec:FMAS}.
Finally, Section~\ref{sec:SRM2D} demonstrates the calculation of a solitary wave for a two-dimensional ($d=2$) model system. This example illustrates how the functionality of {\tt{SWtools}} can be extended by the implementation of further methods, such as a $d=2$ SRM \cite{Musslimani:JOSAB:2004}.

\begin{figure}[!t]
\includegraphics[width=\linewidth]{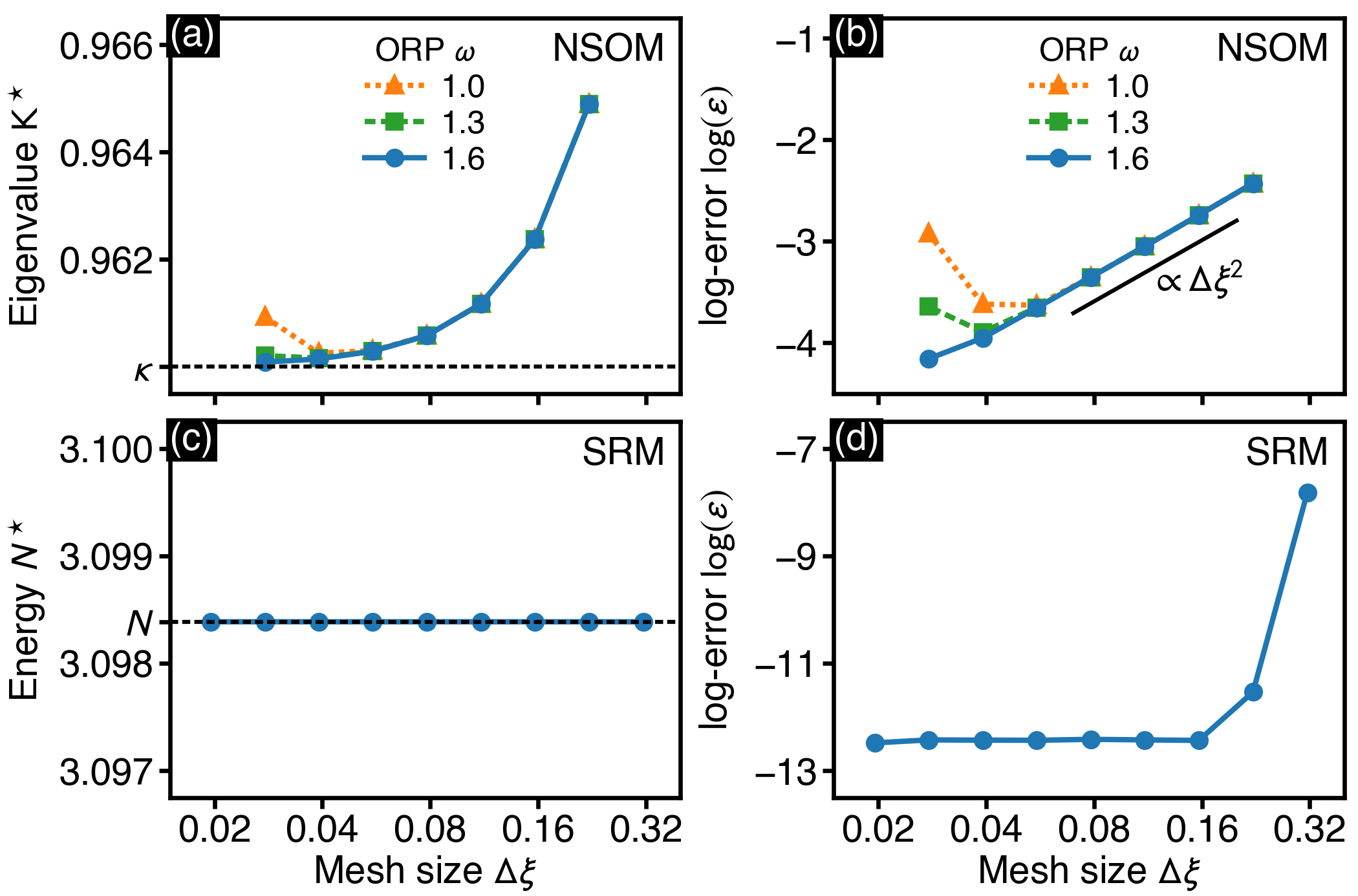}
\caption{
Verification of the solver using an exact soliton solution for the HONSE~(\ref{eq:HONSE_24}).
(a-b) Results of the NSOM for different ORP values $\omega$ and mesh sizes $\Delta \xi$.
(a) Approximation of the soliton eigenvalue $\kappa$.
(b) Relative error~(\ref{eq:eps_glob}).
(c-d) Results of the SRM for different mesh sizes.
(c) Approximation of the soliton energy $N[U]$.
(d) Relative error~(\ref{eq:eps_glob}).
\label{fig:verification}}
\end{figure}

\subsection{Verification of the implementation\label{sec:verification}}
%
%In order to verify the implementation of the provided iterative solvers, 
We consider the complex envelope $\psi\equiv \psi(\eta,\xi)$, governed by the dimensionless HONSE %\cite{Karlsson:OC:1994}
\begin{align}
i\partial_\eta\,\psi =\left(c_2 \partial_\xi^2 - c_4 \partial_\xi^4 \right)\,\psi - |\psi|^2\,\psi, \label{eq:HONSE_24}
\end{align}
with  propagation distance $\eta$ and retarded time $\xi$, modeled after Refs.~\cite{Karlsson:OC:1994,Akhmediev:OC:1994,Piche:OL:1996}.
For parameters $c_2<0$, $c_4<0$, previous analysis of Eq.~(\ref{eq:HONSE_24}) resulted in an exact, fixed-parameter solution $\psi(\eta,\xi) = U(\xi)\,\exp(i\kappa \eta)$ with %\cite{Karlsson:OC:1994}
\begin{align}
U(\xi) = \sqrt{\frac{3}{10}\frac{c_2^2}{|c_4|}} {\rm{sech}}^2\left(\sqrt{\frac{1}{20}\frac{c_2}{c_4}} \xi\right), \label{eq:KH_soliton}
\end{align}
and propagation constant $\kappa= 4 c_2^2 / (25 |c_4|)$ \cite{Karlsson:OC:1994}.
Below we use the exact solution for $c_2=-1/2$ and $c_4=-1/24$ to verify the implementation of the solvers provided by {\tt{SWtools}}.
For the chosen parameters, $U$ is characterized by the soliton energy $N[U]\approx 3.0984$ and $\kappa=0.96$.
%
%Below we use this exact soliton solution to verify the implementation of the solvers provided by {\tt{SWtools}}.
%
%Equation~(\ref{eq:HONSE_24}) comprises an instance of the GNSE~(\ref{eq:GNLS}) with $\eta=z$,  $\xi=\tau$, coefficients $c=\left(0,\,\beta_2/2,\,0,\,\beta_4/24\right)$, and functional $F=\gamma |\psi|^2$. 
%
%We consider $\beta_2=-1$, $\beta_4=-1$, and $\gamma=1$, for which $U^\star$ has soliton energy $E^\star \equiv N[U^\star]\approx 3.0984$, and  $\kappa^\star=0.96$.
%
%
Equation~(\ref{eq:HONSE_24}) comprises an instance of the GNSE~(\ref{eq:GNLS}) with coefficients $c_L=\left(0,\,c_2,\,0,\,c_4\right)$, and functional $F=|\psi|^2$. 
%We solve the corresponding NEVP~(\ref{eq:EVP}) using uniform grids $\tau=(\tau_0, \ldots, \tau_M)$ for different mesh sizes $\Delta \tau$, boundaries $\tau_M=-\tau_0=20$, and fixed accuracy threshold ${\tt{tol}}=10^{-12}$.
%
We solve the corresponding NEVP~(\ref{eq:EVP}) for $\xi\in [-20,20] $ on uniform grids with various mesh sizes $\Delta \xi$, and fixed accuracy threshold ${\tt{tol}}=10^{-12}$.
For a numerical solution $U^\star$, we calculate the relative error with respect to the exact solution as
\begin{align}
\epsilon = \lVert U-U^{\star} \rVert/N. \label{eq:eps_glob}
\end{align}
%
%For a numerical solution $U^\star=(U^\star_0,\ldots, U^\star_M)$, we determine the %root-mean-square (RMS) error with respect to the exact solution as
%\begin{align}
%\epsilon_{\rm{glob}} = \left( \frac{\Delta \tau}{N} \sum_{m=0}^M | U(\tau_m) - %U^\star_{m} |^2\right)^{1/2}. \label{eq:eps_glob}
%\end{align}

We first assess the performance of the NSOM (Section~\ref{sec:alg_01}) for the normalization constraint $N_0=N[U]$, and overrelaxation parameter ${\tt{ORP}}=1.5$.
As evident from Fig.~\ref{fig:verification}(a),  the numerical solution $U^\star$ yields an eigenvalue estimate ${\rm{K}}^\star=H[U^\star]/N_0$ that converges to the exact value of $\kappa$ for decreasing time increments $\Delta \xi$.
The RMS error $\epsilon_{\rm{glob}}$ can be seen to decrease $\propto \Delta \xi^2$ [Fig.~\ref{fig:verification}(b)], in accord with the truncation error of the finite-difference approximation for the highest order derivative in Eq.~(\ref{eq:HONSE_24}).
(If the highest order derivative in $\hat{L}$ is of second order, the RMS error is expected to decrease $\propto \Delta \xi^4$. We verified this using the fundamental soliton for the standard nonlinear Schrödinger equation.)
In case of the SRM, especially high accuracy and excellent approximation of the soliton energy $N[U]$ is evident for all considered mesh sizes [Figs.~\ref{fig:verification}(c,d)].
For interpreting the RMS error in Figs.~\ref{fig:verification}(b,d) it is useful to note that at $\Delta \xi=0.16$, the full-width at half-maximum of the soliton intensity distribution ($\xi_{\rm{FWHM}}\approx 1.56$) is coverd by approximately 10 mesh points.

%\clearpage

%\subsection{Ground-state solutions of one-dimensional Bose-Einstein condensates}
%\subsection{Application of the successive overrelaxation method to find ground-state solutions of 1D BECs}
\subsection{Application of the NSOM to find ground-state solutions for a GPE \label{sec:1DBEC}}
We next demonstrate the calculation of ground-state solutions of one-dimensional BECs, modeled after Ref.~\cite{Bao:SIAM:2004}.
We consider the BEC wavefunction \mbox{$\psi\equiv \psi(\eta,\xi)$}, governed by the dimensionless GPE
\begin{align}
i\partial_\eta\,\psi = -\tfrac{1}{2}\partial_\xi^2\,\psi + V(\xi)\,\psi + \beta|\psi|^2\,\psi, \label{eq:1DGPE}
\end{align}
with  evolution time $\eta$, spatial coordinate $\xi$, 
 harmonic trapping potential $V(\xi)=\frac{1}{2}\xi^2$, and interaction coefficient $\beta>0$.
%modeled after Ref.~\cite{Bao:SIAM:2004}.
%
This yields an instance of the NEVP~(\ref{eq:EVP}) with $c_L=(0,-1/2,0,0)$, $F=-\frac{1}{2}\xi^2 - \beta |\psi|^2$, and normalization constraint $N[\psi]=1$.
%
%Let us note that per Ansatz~(\ref{eq:ansatz}), 
Let us note that $U$ denotes the real-valued ground-state solution of the BEC with chemical potential $\mu=-\kappa$.

\begin{figure}[!t]
\includegraphics[width=\linewidth]{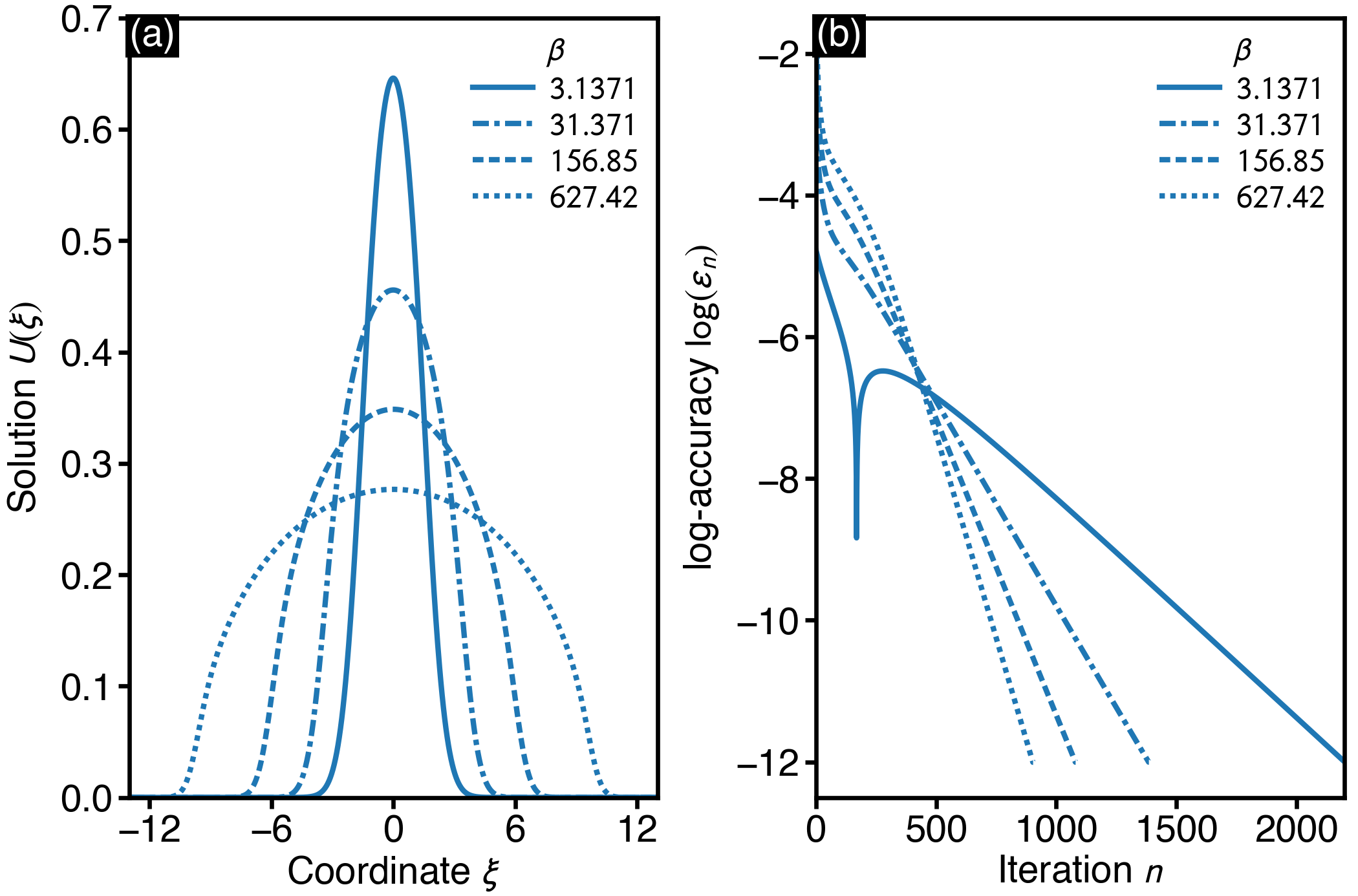}
\caption{
Nonlinear bound states for Eq.~(\ref{eq:1DGPE}) with harmonic oscillator potential, obtained by NSOM.
(a) Solutions $U$ for selected values of the interaction coefficient $\beta$.
(b) Evolution of the accuracy $\epsilon_n$. Iteration is stopped when $\epsilon_n<10^{-12}$.
\label{fig:1DGPE}}
\end{figure}

We solve this constraint nonlinear eigenvalue problem by the NSOM for $\xi\in [-16,16]$, using a uniform grid with mesh size \mbox{$\Delta \xi=1/32$}, and overrelaxation parameter ${\tt{ORP}}=1.5$.
As initial condition we use $U_0(\xi)=\pi^{-1/4}\,e^{-\xi^2/2}$, i.e.\ the exact ground-state in absence of interactions ($\beta=0$). 
To allow for comparison with the results of Ref.~\cite{Bao:SIAM:2004}, we determine the root-mean-square (rms) width of the condensate $\xi_{\rm{rms}}=\int (\xi\,
U)^2~{\rm{d}}\xi$ and its energy $E_\beta[U] = \mu - \frac{\beta}{2} \int U^4~{\rm{d}}\xi$.
A Python script that facilitates this task is detailed in listing~\ref{code:1DGPE}.
For instance, for $\beta=31.371$, we find a solution with peak amplitude $\max(U)\approx 0.4557$, $\xi_{\rm{rms}}\approx 1.6417$, $E_\beta\approx 3.9810$, and $\mu\approx 6.5527$, in excellent agreement with Ref.~\cite{Bao:SIAM:2004}.
Figure~\ref{fig:1DGPE} shows the ground-state solution and the convergence of the accuracy for selected values of $\beta$ (cf.\ Fig.~4.2 of Ref.~\cite{Bao:SIAM:2004}).

\begin{lstlisting}[floatplacement=H, numbers=left, 
captionpos=t, frame=lines,stepnumber=1,numbers=left, numbersep=5pt, xleftmargin=\parindent,
language=Python, 
caption={
Python code for solving the NEVP for Eq.~(\ref{eq:1DGPE}) via NSOM.}, label=code:1DGPE]
import numpy as np
from SWtools import NSOM

# -- SETUP AND INITIALIZATION 
xi = np.linspace(-16, 16, 1025)
# ... NEVP INGREDIENTS 
cL = (0, -0.5, 0, 0)
beta = 31.371
F = lambda I, xi: -(0.5*xi**2 + beta*I)
N0 = 1.0
# ... INITIAL CONDITION 
U0 = np.exp(-xi**2/2)/np.pi**0.25
# ... NSOM INSTANTIATION
myS = NSOM(xi, cL, F, ORP=1.5)

# -- NSOM SOLUTION PROCEDURE
myS.solve(U0, N0)

# -- POSTPROCESSING (RMS WIDTH AND ENERGY)
U, mu = myS.U, -myS.kap
xi_rms = np.sqrt(np.trapz((xi*U)**2,x=xi))
E = mu - 0.5*beta*np.trapz(U**4,x=xi)
print(f"max(U) = {np.max(U):5.4F}")
print(f"xi_rms = {xi_rms:5.4F}")
print(f"E_beta = {E:5.4F}")
print(f"mu     = {mu:5.4F}")
\end{lstlisting}

%\clearpage

\subsection{Application of the SRM to find solutions of a HONSE\label{sec:GQS}}

We next demonstrate the calculation of ``traveling'' soliton solutions in nonlinear optical media, modeled after Refs.~\cite{Tsoy:PRA:2024,Melchert:PRA:2024}. 
We consider the slowly varying envelope $\psi\equiv \psi(\eta,\xi)$, governed by the dimensionless HONSE
\begin{align}
i\partial_\eta\,\psi =\left(ic_1 \partial_\xi + c_2 \partial_\xi^2 + i c_3 \partial_\xi^3 - c_4\partial_\xi^4 \right)\,\psi - |\psi|^2\,\psi, \label{eq:HONSE}
\end{align}
with  propagation distance $\eta$ and retarded time $\xi$.
%for paramters that satisfy $c_2<0$, $c_4<0$, and $8 c_2 c_4 > 3 c_3^2$,
%modeled after Ref.~\cite{Tsoy:PRA:2024},
%with  propagation distance $z$, retarded time $\tau$, dispersion parameters $(\beta_2,\beta_3,\beta_4)=(-0.2,0.2,-1)$, and nonlinear coefficient $\gamma=1$. 
%
%The parameters $\beta_2$, $\beta_3$, and $\beta_4$ specify second-, third-, and fourth-order dispersion, respectively. 
%
%To seek soliton solutions for Eq.~(\ref{eq:HONSE}) with nonzero soliton velocity $c_1=-1/v_S$, we make use of the parameter $c_1$ in the linear operator~(\ref{eq:L}), allowing to define the retarded time $\tau^\prime = \tau - c_1 z$ with $c_1=-1/v_S$. 
Here, we seek such traveling solitons as stationary solutions in a reference frame moving at nonzero inverse velocity $c_1$.
Equation~(\ref{eq:HONSE}) is nonintegrable and admits no general analytical soliton solution.
Depending on the choice of parameters, its solitons can exhibit oscillating tails \cite{Akhmediev:OC:1994,deSterke:OC:2023,Melchert:PRA:2024}. 
Secifically, we consider the NEVP~(\ref{eq:EVP}) for $c_L=(-0.07,\, -0.1,\, 0.0333,\, -0.0417)$, and $F= |\psi|^2$.
%, modeled after Ref.~\cite{Tsoy:PRA:2024}.
%, and allows to complement parts of the simulations reported in Ref.~\cite{Tsoy:PRA:2024}.

\begin{figure}[!t]
\includegraphics[width=\linewidth]{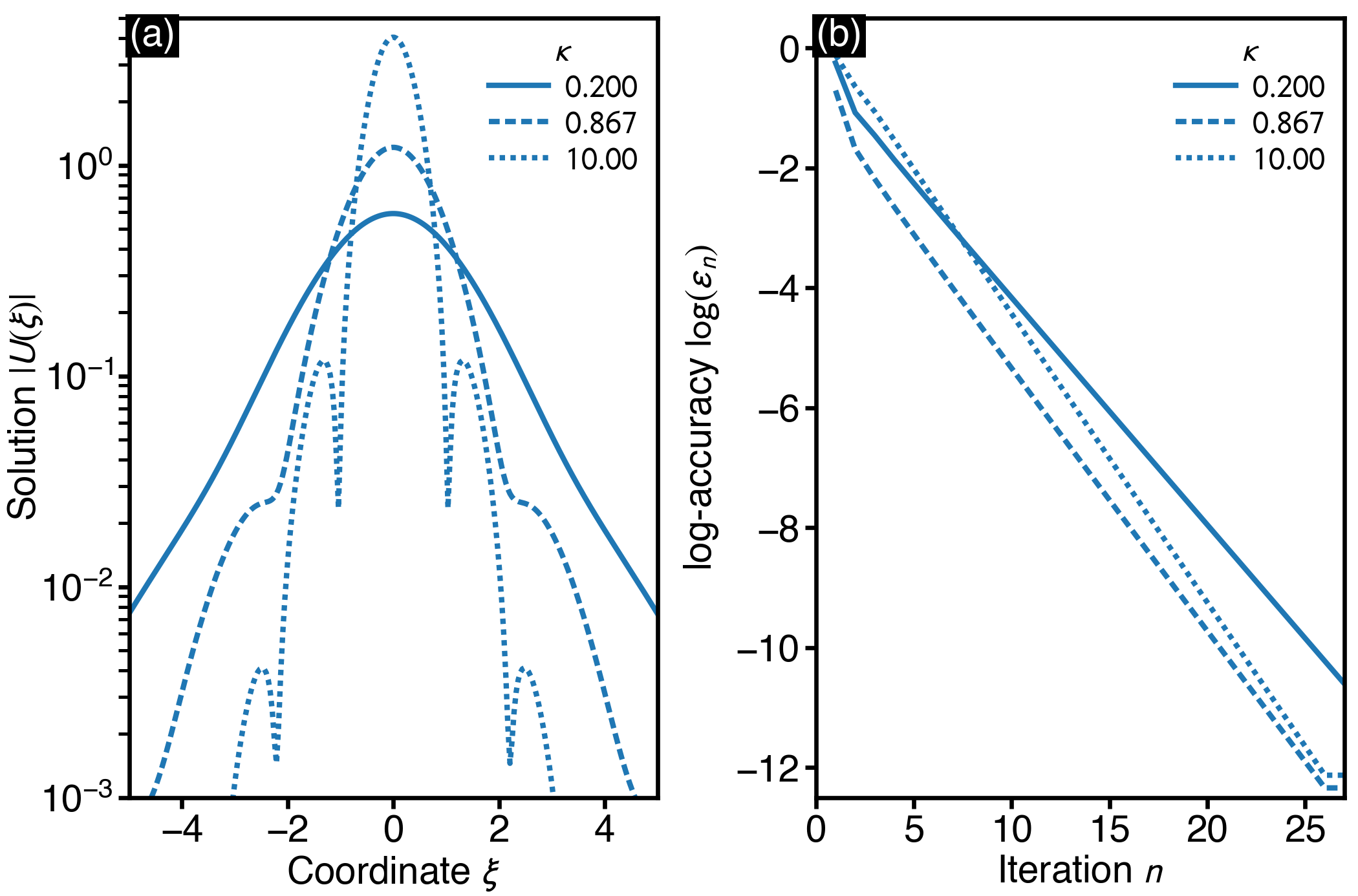}
\caption{
Soliton solutions for Eq.~(\ref{eq:HONSE}) obtained by the SRM.
(a) Solutions $U$ for selected values of the propagation constant $\kappa$ and fixed reference frame velocity $c_1=-0.07$.
(b) Evolution of the accuracy $\epsilon_n$. Iteration is stopped when $\epsilon_n<10^{-12}$.
\label{fig:HONSE}}
\end{figure}

We solve the NEVP in the interval $\xi\in [-20,20]$, using a uniform grid with $2^{10}$ mesh points and initial condition $U_0(\xi)=e^{-\xi^2}$.
%
%We solve Eq.~(\ref{eq:EVP}) for $(\beta_2,\beta_3,\beta_4)=(-0.2,0.2,-1)$ and $\gamma=1$ using the SRM of Sect.~\ref{sec:alg_02} for $\tau^\prime \in [-20,20]$, using a discretization with $2^{10}$ equidistant mesh points and default error tolerance (${\tt tol}=10^{-12}$).
%
%As initial condition we use $U_0(\tau^\prime)=e^{-\tau^{\prime 2}}$. 
%
A Python script that facilitates this task is detailed in listing~\ref{code:HONSE}.
At a given iteration step $n$, the accuracy is calculated as $\epsilon_n = \max(|U^{(n-1)}-U^{(n)}|)$ (lines 16\,ff.\ of listing~\ref{code:HONSE}).
To facilitate comparison with the results of Ref.~\cite{Tsoy:PRA:2024}, we determine the soliton center frequency \mbox{$k_c = \int k\,|u(k)|^2~{\rm{d}}k/\int |u(k)|^2~{\rm{d}}k$}, where $u(k)$ is the Fourier transform of the solution $U(\xi)$ (line 24\,f.\ of listing~\ref{code:HONSE}).
In order to respect the sign choices and normalization of the transforms~(\ref{eq:FT}), we use the numpy-native function {\tt{ifft}} to implement Eq.~(\ref{eq:FT_FT}).
For $\kappa=0.867$ we find a solution with peak amplitude $\max(|U|)\approx 1.215$, $N[U]\approx 2.001$, and $k_c\approx 0$. 
It reproduces one of the solitons obtained earlier by exploring pulse propagation simulations \cite{Tsoy:PRA:2024}. 
Results for selected values of $\kappa$ are shown in Fig.~\ref{fig:HONSE}.
%Figure~\ref{fig:HONSE} shows solutions and the convergence of the accuracy for selected wavenumbers $\kappa$.
%and fixed soliton velocity $1/v_s=0.07$. 
%

\begin{lstlisting}[floatplacement=H, numbers=left, 
captionpos=t, frame=lines,stepnumber=1,numbers=left, numbersep=5pt, xleftmargin=\parindent,
language=Python, 
caption={Python code for solving the NEVP for Eq.~(\ref{eq:HONSE}) via SRM.}, label=code:HONSE]
import numpy as np
from numpy.fft import fftfreq, ifft
from SWtools import SRM

# -- SETUP AND INITIALIZATION 
xi = np.linspace(-20, 20, 2**10)
# ... NEVP INGREDIENTS 
cL = (-0.07, -0.2/2, 0.2/6, -1./24)
F = lambda I, xi: I
kap = 0.867
# ... INITIAL CONDITION 
U0 = np.exp(-xi**2)
# ... SRM INSTANTIATION
myS = SRM(xi, cL, F)
# ... CUSTOM ACCURACY
acc = lambda xi, U, V: np.max(np.abs(U-V))

# -- SRM SOLUTION PROCEDURE 
myS.solve(U0, kap, acc_fun = acc)

# -- POSTPROCESSING
U, N = myS.U, myS.N
k = fftfreq(xi.size, d=xi[1]-xi[0])*2*np.pi
Ik = np.abs(ifft(U))**2
kc = np.trapz(k*Ik, x=k)/np.trapz(Ik, x=k)
print(f"max(U) = {np.max(np.abs(U)):5.4F}")
print(f"N[U]   = {N:5.4F}")
print(f"kc     = {kc:5.4F}")
\end{lstlisting}

%\clearpage

\subsection{Application of the SRM to find solutions of a GNSE with periodic nonlinearity  \label{sec:nlinMicro}}

%In this section we demonstrate the calculation of nonlinear bound states for a GNSE that models the propagation of laser beams in a medium where the nonlinear refractive index is modulated along the transverse direction \cite{Fibich:PD:2006}.
%
Below we demonstrate the calculation of nonlinear bound states in a medium where the nonlinear refractive index is modulated along the transverse direction, modeled after Ref.~\cite{Fibich:PD:2006}.
We consider the slowly varying envelope $\psi=\psi(\eta,\xi)$, goverened by the dimensionless GNSE
\begin{align}
i\partial_\eta \,\psi = -\partial_\xi^2\,\psi - [1+m(\xi)] \,|\psi|^2 \psi, \label{eq:1DNSE}
\end{align}
with propagation distance $\eta$, transverse coordinate $\xi$, and periodic nonlinear microstructure $m(\xi) = \alpha \cos(4\pi \xi)$.
%
%The beam is centered at the local maximum (minimum) of the microstructure for $\alpha>0$ ($\alpha<0$).
%
This yields an instance of the NEVP~(\ref{eq:EVP}) with $c_L=(0,-1,0,0)$, and $F=[1+m(\xi)]\,|\psi|^2$, for which we consider $\alpha<0$ and $\kappa=1$.
%
%Below we reproduce parts of the simulations reported in Ref.~\cite{Fibich:PD:2006} for $\kappa = 1$ and selected values of $\alpha<0$.

\begin{figure}[!t]
\includegraphics[width=\linewidth]{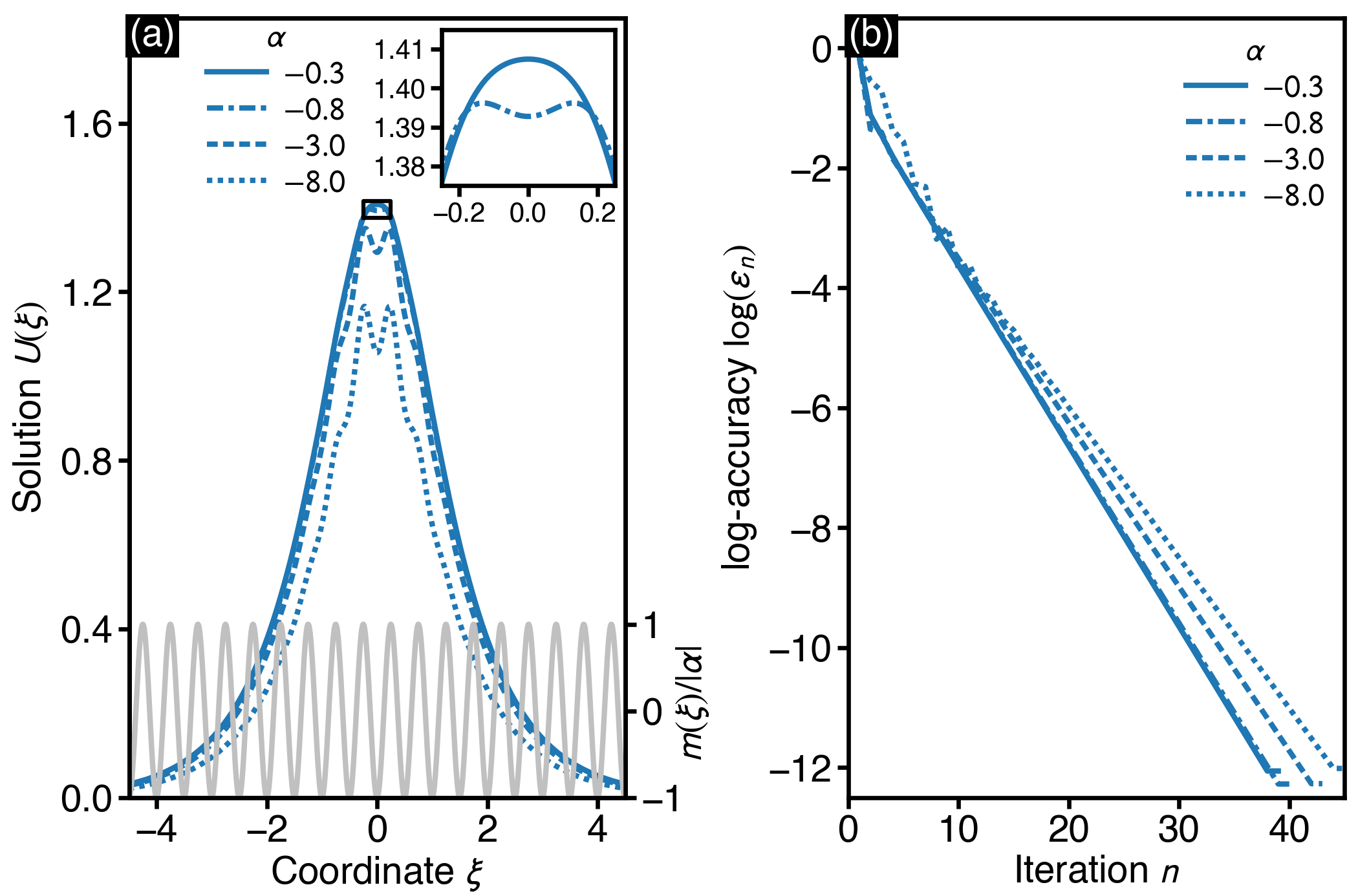}
\caption{
Nonlinear bound states for Eq.~(\ref{eq:1DNSE}) obtained by the SRM.
(a) Solutions $U$ for selected microstructure parameters $\alpha$. The inset shows a close-up view of the peak region for $\alpha=-0.3$ and $-0.8$. Gray line shows $m(\xi)/|\alpha|$. 
(b) Evolution of the accuracy $\epsilon_n$. Iteration is stopped when 
$\epsilon_n<10^{-12}$.
\label{fig:nlinMicro}}
\end{figure}

To solve the NEVP in the interval $\xi\in [-20,20]$, we use a uniform grid with $2^{12}$ mesh points, and initial condition $U_0(\xi)=e^{-\xi^2}$ centered at a local minimum of $m(\xi)$.
A Python script that facilitates this task is detailed in listing~\ref{code:nlinMicro}.
Results for selected values of $\alpha$ are shown in Fig.~(\ref{fig:nlinMicro}) (cf.\ Fig.~4 of Ref.~\cite{Fibich:PD:2006}).
Specifically, for $\alpha=-0.8$ the solution has the local minimum $U(0)\approx 1.393$, and two adjacent global maxima $\max(U)\approx 1.396$.

\begin{lstlisting}[floatplacement=H, numbers=left, 
captionpos=t, frame=lines,stepnumber=1,numbers=left, numbersep=5pt, xleftmargin=\parindent,
language=Python, 
caption={Python code for solving the NEVP for Eq.~(\ref{eq:1DNSE}) via SRM.}, label=code:nlinMicro]
import numpy as np
from SWtools import SRM

# -- SETUP AND INITIALIZATION 
xi = np.linspace(-20, 20, 2**12)
# ... NEVP INGREDIENTS 
cL = (0, -1., 0, 0)
m = lambda xi: -0.8*np.cos(4*np.pi*xi)
F = lambda I, xi: (1 + m(xi))*I
kap = 1.0
# ... INITIAL CONDITION
U0 = np.exp(-xi**2)
# ... SRM INSTANTIATION
myS = SRM(xi, cL, F)

# -- SRM SOLUTION PROCEDURE
myS.solve(U0, kap)

# -- POSTPROCESSING
U, N = myS.U.real, myS.N
print(f"# N      = {N:4.3F}")
print(f"# max(U) = {np.max(U):4.3F}")
print(f"# U(0)   = {U[xi.size//2]:4.3F}")
\end{lstlisting}

%\clearpage

\subsection{Application of the NSOM to find excited states for a model GPE \label{sec:excitedStates}}
To demonstrate the calculation of nonlinear excited-states, we construct a simple problem based on a quantum harmonic oscillator.
We consider a wavefunction \mbox{$\psi\equiv \psi(\eta,\xi)$}, governed by the dimensionless GPE
\begin{align}
i\partial_\eta\,\psi = -\tfrac{1}{2}\partial_\xi^2\,\psi + \left(\tfrac{1}{2}\xi^2 + \beta |\psi|^2\right)\,\psi, 
\label{eq:1DGPE_2}
\end{align}
with  evolution time $\eta$, spatial coordinate $\xi$, harmonic trapping potential, and interaction coefficient $\beta=1$. 
This yields an instance of the NEVP~(\ref{eq:EVP}) with $c_L=(0,-1/2,0,0)$, $F=-\frac{1}{2}\xi^2 -|\psi|^2$, and normalization constraint $N[\psi]=1$. 
Ansatz~(\ref{eq:ansatz}) then introduces a real-valued function $U$ for the chemical potential $\mu=-\kappa$.

\begin{figure}[!t]
\includegraphics[width=\linewidth]{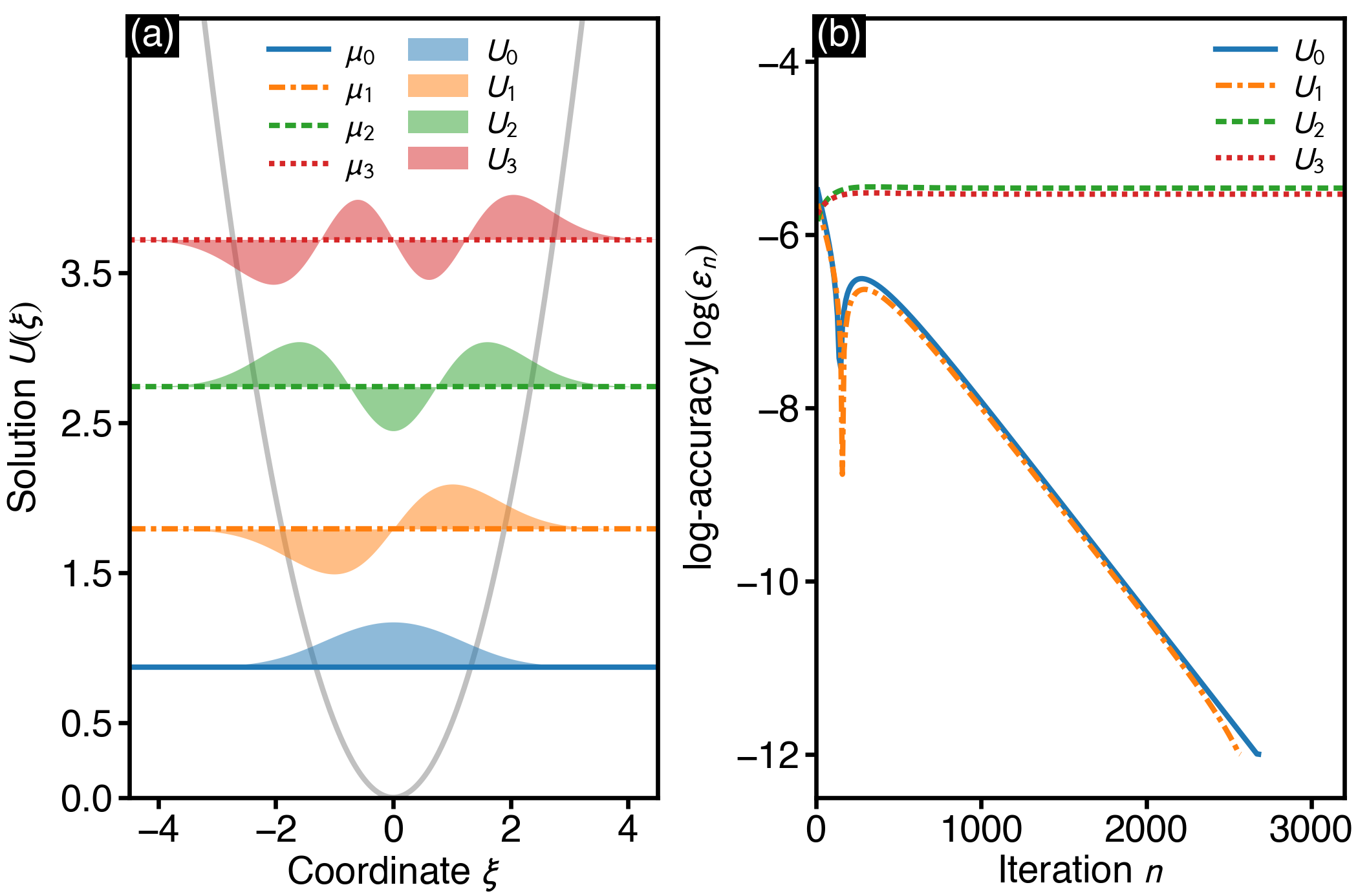}
\caption{
Nonlinear bound states for Eq.~(\ref{eq:1DGPE_2}) obtained by the NSOM.
(a) Ground state solution $U_0$ and first three exited states $U_1, \ldots, U_3$, indicated at their respective chemical potential $\mu$. 
Gray solid line indicates the harmonic trapping potential.
(b) Evolution of the accuracy $\epsilon_n$. Iteration is stopped when $\epsilon_n<10^{-12}$, or when the maximum number of iterations (${\tt{maxiter}}=10^4$) is exceeded.
\label{fig:excitedStates}}
\end{figure}

We solve this constraint nonlinear eigenvalue problem by the NSOM for $\xi\in [-8,8]$, using a discretization with mesh size \mbox{$\Delta \xi=1/32$}, and overrelaxation parameter ${\tt{ORP}}=1.5$.
A Python script that facilitates this task is detailed in listing~\ref{code:excitedStates}.
The ground-state and first three excited-states are obtained one by one using the orthogonalization method implemented with {\tt{SWtools}}.
Therefore, the class method {\tt{solve()}} accepts the keyword argument {\tt{ortho\_set}} (see \ref{sec:NSOM_subclass}), allowing to specify a list of previously found solutions (see lines 24, 28, and 32 of listing~\ref{code:excitedStates}).
During the iteration process, any overlap with these solutions is eliminated.
As initial conditions we use the exact eigenfunctions of the quantum harmonic oscillator ($\beta=0$). 
Figure~\ref{fig:excitedStates} shows the obtained solutions and their convergence properties.
For the ground state ($U_0$) and first excited state ($U_1$), the method is highly accurate. 
The reduced accuracy in case of higher order excited states, demonstrated by the saturation of the accuracy at approx.\ $10^{-6}$, 
%has previously been reported for similar orthonormalization schemes \cite{Lehtovaara:JCP:2007}.
%
lies in the efficiency of the NSOM to drive a maintained solution away from an excited state whenever a lower-lying state of the same parity is available.
We carefully checked that the exact results for the quantum harmonic oscillator are met in the limit $\beta\to 0$.
%
%\paragraph{Technical note}
%
As a technical note, let us point out that the Gauss-Seidel update implemented with the NSOM does not preserve the symmetry of a trial function.
Therefore, without the orthogonalization approach, an odd-parity initial condition does not automatically converge to the first excited state.

\begin{lstlisting}[floatplacement=H, numbers=left, 
captionpos=t, frame=lines,stepnumber=1,numbers=left, numbersep=5pt, xleftmargin=\parindent,
language=Python, 
caption={Python code for solving the NEVP for Eq.~(\ref{eq:1DGPE_2}) via NSOM.}, label=code:excitedStates]
import numpy as np
from SWtools import NSOM

# -- SETUP AND INITIALIZATION 
xi = np.linspace(-8, 8, 512)
# ... NEVP INGREDIENTS 
cL = (0,-0.5,0,0)
F = lambda I, xi: -0.5*xi**2 - I
N0 = 1.0
# ... INITIAL CONDITIONS 
UI0 = np.exp(-xi**2/2)/np.pi**0.25
UI1 = 2*xi*UI0/np.sqrt(2)
UI2 = (4*xi**2-2)*UI0/np.sqrt(8)
UI3 = (8*xi**3-12*xi)*UI0/np.sqrt(48)
# ... INSTANTIATE NSOM
myS = NSOM(xi, cL, F, ORP=1.5)

# -- NSOM SOLUTION PROCEDURE
# ... GROUND STATE
myS.solve(UI0, N0)
U0, mu0 = myS.U, -myS.kap
# ... 1ST EXCITED STATE
myS.solve(UI1, N0, ortho_set=[U0])
U1, mu1 = myS.U, -myS.kap
# ... 2ND EXCITED STATE
myS.solve(UI2, N0, ortho_set=[U0, U1])
U2, mu2 = myS.U, -myS.kap
# ... 3RD EXCITED STATE
myS.solve(UI3, N0, ortho_set=[U0, U1, U2])
U3, mu3 = myS.U, -myS.kap
\end{lstlisting}

%\clearpage

\subsection{Software integration with {\tt{py-fmas}}\label{sec:FMAS}}

\begin{figure}[!t]
\includegraphics[width=\linewidth]{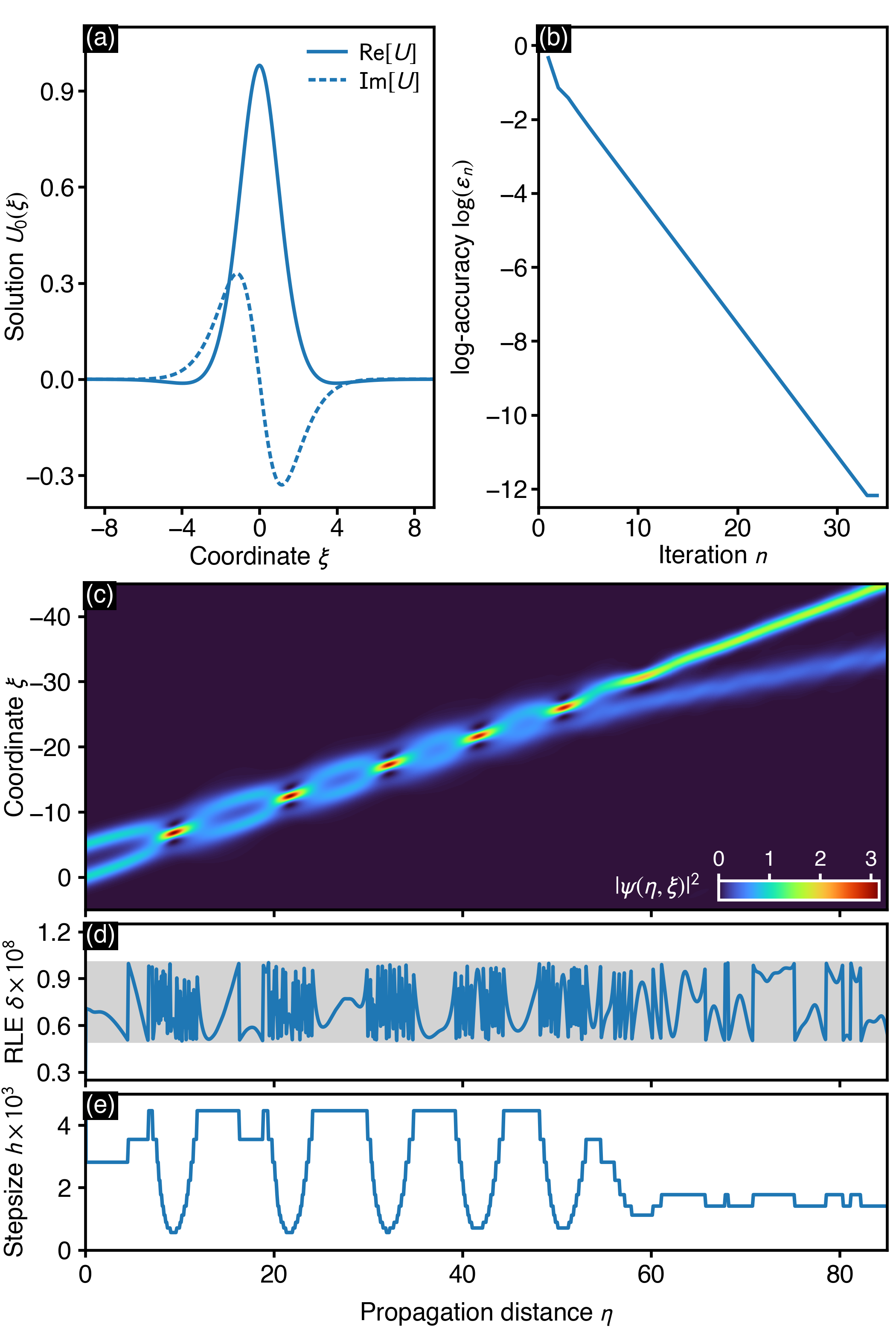}
\caption{
%
%Example combining {\tt{SWtools}} and {\tt{py-fmas}} to study solutions of the HONSE~(\ref{eq:HONSE}).
%
Combining {\tt{SWtools}} and {\tt{py-fmas}} to study the interaction of solitons of Eq.~(\ref{eq:HONSE}).
(a-b) Solution of the NEVP using the SRM implemented in {\tt{SWtools}}.
(a) Complex-valued soliton $U_0$ for $\kappa=0.681$.
(b) Accuracy $\epsilon_n$. Iteration is stopped when $\epsilon_n<10^{-12}$.
(c-e) Solution of the initial value problem using an adaptive stepsize method implemented in {\tt{py-fmas}}.
(c) Propagation dynamics of the intensity $|\psi|^2$.
(d) Variation of the relative local error (RLE) $\delta$.
(e) Variation of the stepsize $h$.
\label{fig:FMAS}}
\end{figure}

Below we demonstrate how {\tt{SWtools}} can be used in conjunction with {\tt{py-fmas}}, allowing to study the interaction dynamics of solitons governed by the HONSE~(\ref{eq:HONSE}).
A short workflow that reproduces one of the propagation scenarios  of Ref.~\cite{Melchert:PRA:2024} for parameters $c_L=(0,-1/2,1/12,-1/24)$ and $F=|\psi|^2$ is shown in listing~\ref{code:FMAS}.
First, in lines 14--15, a traveling soliton $U_0(\xi)$ with inverse group-velocity $1/v_0=-0.458$ and propagation constant $\kappa=0.618$ is obtained by solving the NEVP using the SRM implemented in {\tt{SWtools}}.
The resulting complex-valued solution is shown in Figs.~\ref{fig:FMAS}(a,b).
In line 18, this soliton is used to compose the initial condition $\psi(0,\xi)= U_0(\xi)+U_0(\xi-\Delta \xi)\,e^{i\Delta \phi}$, consisting of two identical solitons with separation $\Delta \xi=-5$ and phase mismatch $\Delta \phi \approx 2.5$. 
Subsequently, in lines 18--25, the initial value problem for Eq.~(\ref{eq:HONSE}) is solved using the adaptive stepsize local error method (LEM) \cite{Sinkin:JLT:2003}, implemented in {\tt{py-fmas}} \cite{Melchert:CPC:2022}.
The LEM keeps the relative local error $\delta$, estimated by step-doubling and local extrapolation \cite{Press:BOOK:2007}, within the target range $(\delta_G/2, \delta_G)$ with goal local error $\delta_G=10^{-8}$ specified in line 21.
After termination of the LEM, the results are stored in 
numpy-native {\tt{npz}}-fomat in lines 28--37.
The result of the pulse propagation simulation is shown in Fig.~\ref{fig:FMAS}(c), indicating the formation of a two-pulse bound state that decays after five collisions \cite{Melchert:PRA:2024}.
As evident from Figs.~\ref{fig:FMAS}(d,e), guided by the relative local error $\delta$, the stepsize $h$ decreases when the pulses engage in a collision and it increases in between successive collisions. 

\begin{lstlisting}[floatplacement=H, numbers=left, 
captionpos=t, frame=lines,stepnumber=1,numbers=left, numbersep=5pt, xleftmargin=\parindent,
language=Python, 
caption={Python code showing how to combine {\tt{SWtools}} and {\tt{py-fmas}} \cite{Melchert:CPC:2022} to study the interaction of solitons of Eq.~(\ref{eq:HONSE}).}, label=code:FMAS]
import numpy as np
from SWtools import SRM, FT, IFT, FTFREQ
from fmas.solver import LEM

# -- SET UP DOMAIN AND MODEL 
xi = np.linspace(-50, 50, 2**13)
k = FTFREQ(xi.size, d=xi[1]-xi[0])*2*np.pi
c1, c2, c3, c4 = 0.458, -1/2, 1/12, -1/24
F = lambda I, xi: I
N = lambda U: F(np.abs(U)**2,xi)*U
kap = 0.618

# -- SET UP AND SOLVE NEVP 
NEVP = SRM(xi, (c1,c2,c3,c4), F)
NEVP.solve(np.exp(-xi**2), kap)

# -- SET UP AND SOLVE IVP
u0 = FT(NEVP.U)*(1+np.exp(-5j*k+2.49826j))
Lk = 1j*(c2*k**2 + c3*k**3 + c4*k**4)
Nk = lambda u: 1j*FT(N(IFT(u)))
IVP = LEM(Lk, Nk, del_G=1e-8)
IVP.set_initial_condition(k, u0)
IVP.propagate(z_range = 90.,
              n_steps = 1000,
              n_skip = 1)

# -- SAVE DATA IN NPZ-FORMAT
res = { 
  'xi':  xi,
  'U0':  NEVP.U,
  'it':  NEVP.iter_list,
  'acc': NEVP.acc_list,
  'eta': IVP.z,
  'U':   IVP.utz,
  'del': IVP._del_rle,
  'h':   IVP._dz_a }        
np.savez_compressed('res.npz',**res)
\end{lstlisting}

\subsection{Extension of the SRM to $d=2$\label{sec:SRM2D}}

In addition to the one-dimensional examples discussed in the main text, we here present an example for a $d=2$ NSE with transverse vector $\vec{\xi}=(\xi_1,\xi_2)$, modeled after Refs.~\cite{Yang:OL:2003,Ostrovskaya:OE:2004,Yang:SIAM:2008}. 
We consider the envelope \mbox{$\psi\equiv \psi(\eta,\xi)$}, governed by the dimensionless NSE
\begin{align}
i\partial_\eta\,\psi = -\partial_{\xi_1}^2\,\psi - \partial_{\xi_2}^2\,\psi - V(\vec{\xi})\,\psi - |\psi|^2\,\psi, \label{eq:2DNSE}
\end{align}
with  propagation distance $\eta$, transverse coordinates $\xi_{1,2}$, and periodic potential $V(\vec{\xi})=V_0\,[\cos^2(\xi_1)+\cos^2(\xi_2)]$.
This represents an instance of the NEVP~(\ref{eq:EVP}) with $\hat{L}=(i\partial_{\xi_1})^2 + (i\partial_{\xi_2})^2$ and $F=V+|\psi|^2$.
To solve the bare NEVP for Eq.~(\ref{eq:2DNSE}), we provide the extension module {\tt{SWtools\_ext\_SRM2D}}, documented online \cite{SWtools:GH:2025}, which extends the functionality of {\tt{SWtools}} by implementing a custom SRM for $d=2$.

\begin{figure}[!t]
\includegraphics[width=\linewidth]{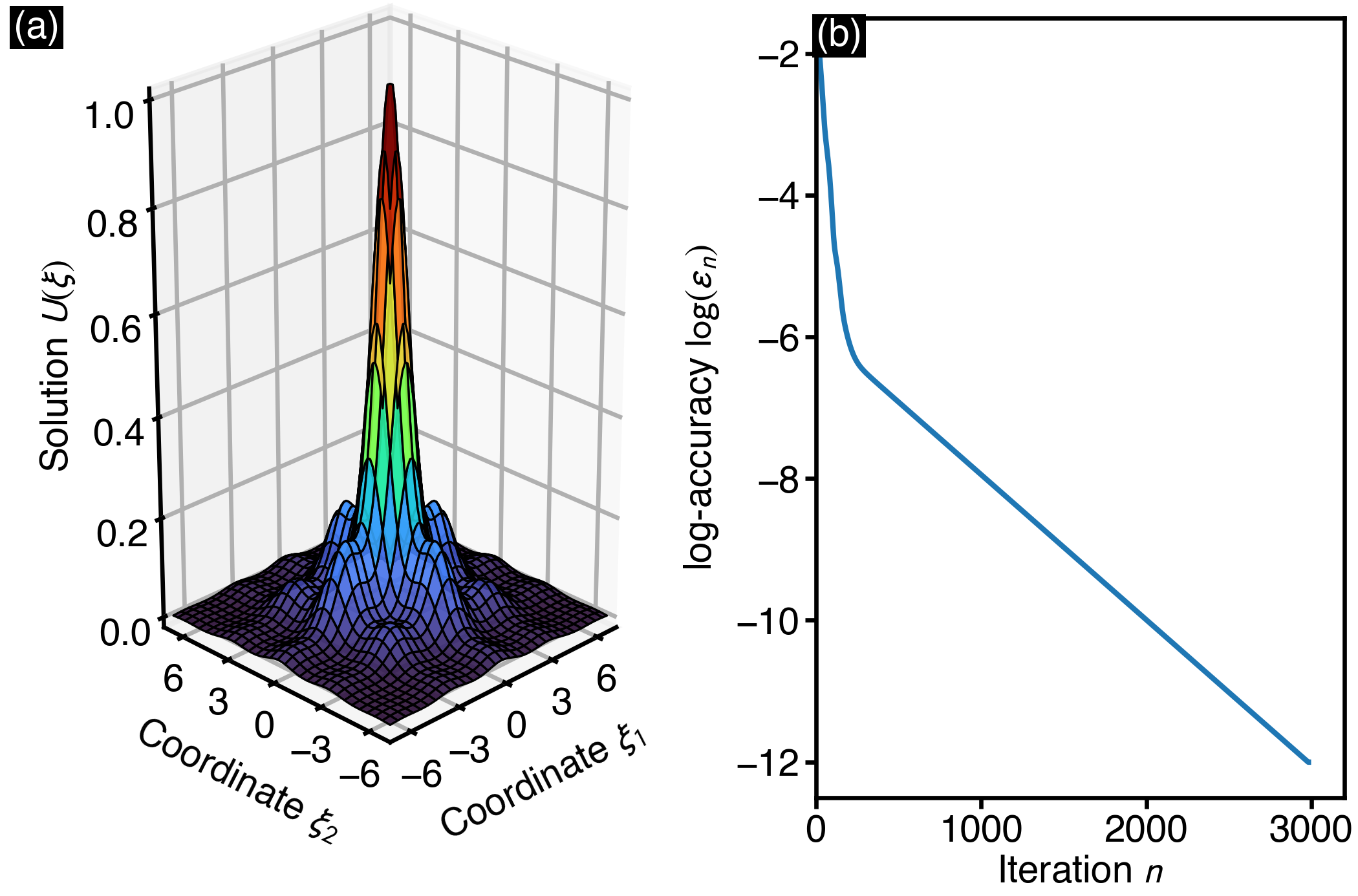}
\caption{
Nonlinear bound state of Eq.~(\ref{eq:1DGPE}), obtained by the $d=2$ SRM.
(a) Solution $U(\xi)$, with $\vec{\xi}=(\xi_1,\xi_2)$, for potential height $V_0=3$ and propagation constant $\kappa=3.7045$.
(b) Evolution of the accuracy $\epsilon_n$. Iteration is stopped when $\epsilon_n<10^{-12}$.
\label{fig:2DNSE}}
\end{figure}

To solve the NEVP using the $d=2$ SRM in the domain $\vec{\xi}\in [-20,20]\times [-20,20]$, we use a uniform grid with $2^8 \times 2^8$ mesh points, and initial condition $U_0(\vec{\xi})=e^{-(\xi_1^2+\xi_2^2)}$.
A Python script that facilitates this task is detailed in listing~\ref{code:2DSRM}.
Results for the potential height $V_0=3$ and propagation constant $\kappa=3.7045$ are shown in Fig.~(\ref{fig:2DNSE}).
The corresponding solution has peak-amplitude $\max(U)\approx1.03$, and exhibits $N[U]\approx 3.00$, and $H[U]\approx11.11$.

\begin{lstlisting}[floatplacement=H, numbers=left, 
captionpos=t, frame=lines,stepnumber=1,numbers=left, numbersep=5pt, xleftmargin=\parindent,
language=Python, 
caption={Python code for solving the NEVP for Eq.~(\ref{eq:2DNSE}) using the $d=2$ SRM.}, label=code:2DSRM]
import numpy as np
from SWtools_SRM2D import SRM2D

# -- SETUP AND INITIALIZATION 
x1 = np.linspace(-20, 20, 2**8)
x2 = np.linspace(-20, 20, 2**8)
X1, X2 = np.meshgrid(x1, x2, indexing='ij')
# ... NEVP INGREDIENTS 
cL = (-1., -1.)
V0=3.
cos2 = lambda x: np.cos(x)**2
V = lambda x: V0*(cos2(x[0]) + cos2(x[1]))
F = lambda I, xi: V(xi) + I
kap = 3.7045
# ... INITIAL CONDITION 
U0 = np.exp(-(X1**2+X2**2))
# ... SRM INSTANTIATION
myS = SRM2D((X1, X2), cL, F)

# -- 2DSRM SOLUTION PROCEDURE
myS.solve(U0, kap)

# -- POSTPROCESSING
U, H, N = myS.U, myS.H, myS.N
print(f"max(U) = {np.max(U):5.4F}")
print(f"H[U]   = {H:5.4F}")
print(f"N[U]   = {N:5.4F}")
\end{lstlisting}

\section{Discussion\label{sec:impact}}

\paragraph{Previous usage}
{\tt{SWtools}} is derived from our research software and has already successfully contributed to the process of scientific discovery in the field of nonlinear optics. 
It has allowed us to calculate numerically exact soliton solutions for GNSE type models in parameter ranges that where previously not accessible \cite{Melchert:PRA:2024,Melchert:arxiv:2024}.
%
%{\tt{SWtools}} is derived from our research software and has successfully contributed to the process of scientific discovery in the field of nonlinear optics  \cite{Melchert:PRA:2024,Melchert:arxiv:2024}.
%
In these previous works we considered Eq.~(\ref{eq:GNLS}) with parameters $c_L=(0,c_2, c_3,c_4)$, describing the dispersion characteristics of the optical medium, and $F=\gamma |\psi|^2$, where $\gamma$ specifies the strength of self-phase modulation. 
In Ref.~\cite{Melchert:PRA:2024} we used the presented tools to determine SW solutions for $c_2<0$, $c_4<0$, and $8 c_2 c_4 > 3 c_3^2$, specifying a model with anomalous dispersion in the entire frequency domain.
The obtained solutions where essential for our systematic study of inelastic collisions that demonstrated the formation of short-lived two-pulse bound states. 
Let us emphasize that {\tt{SWtools}} allowed to directly calculate and characterize these solutions, complementing previous studies in which localized solutions were obtained only indirectly by exploring pulse propagation simulations~\cite{Tsoy:PRA:2024}.
%
%In addition, we were able to complement previous findings~\cite{Tsoy:PRA:2024}, where localized solutions were obtained only indirectly by exploring pulse propagation simulations.
%
%{\tt{SWtools}} allowed to directly calculate and characterize these solutions reliably. 
%
In Ref.~\cite{Melchert:arxiv:2024} we employed the presented tools to determine SW solutions for a related, yet very different GNSE with $c_2>0$, $c_3=0$, and $c_4<0$, exhibiting alternate domains of normal and anomalous dispersion \cite{Melchert:PRL:2019,Melchert:OPTIK:2023}.
This model has only recently begun to attract attention in the field of nonlinear optics \cite{Melchert:PRL:2019,Tam:PRA:2020,Tsoy:PRA:2024}, and is still essentially unexplored.
%
%As a special feature it exhibits alternate domains of normal and anomalous dispersion \cite{Melchert:PRL:2019,Melchert:OPTIK:2023}.
%
We were able to demonstrate special SW solutions which provide properties that are highly demanded in ultrafast science. 
They display carrier-envelop-phase stability and are characterized by spectra that can extend over extremely wide frequency-ranges, encompassing a domain of normal dispersion in which solitons ordinarily cannot persist. They further exhibit unusual energy scaling, are robust against nonlinear perturbations and independent of fine details of the dispersion profile. 
It further allowed us to numerically substantiate the existence of nonlinear photonic meta-atoms, i.e.\ simplified solutions for the weakly nonlinear limit of systems of GNSEs \cite{Melchert:OL:2023,Melchert:OPTIK:2023}.

\paragraph{Pursuit of research questions}
{\tt{SWtools}} makes it affordable and convenient to perform extensive parameter sweeps for the characterization of families of solitary waves $\kappa \to U(\xi; \kappa)$, which emerges as a current topic in the field of nonlinear optics \cite{Melchert:OL:2023,Tam:PRA:2020,Melchert:PRA:2024,Tsoy:PRA:2024}.
This permits to direcly address questions such as ``How do the functionals $N[U]$ and $H[U]$ of a soliton solution $U$ depend on its propagation constant $\kappa$?'', helping to clarity whether the obtained solutions are stable or instable \cite{Rose:PD:1988,Weinstein:BOOK:2015}. 
%
%(The solution is orbitally stable if $\partial_\kappa N[U(\xi;\kappa)] > 0$~\cite{Rose:PD:1988}.)
%
Let us emphasize that {\tt{SWtools}}  can be used on its own, or as extension module for {\tt{GNLStools}} \cite{Melchert:SFX:2022} and {\tt{py-fmas}} \cite{Melchert:CPC:2022}, allowing to study the propagation dynamics of the obtained solutions in terms of the initial-value problem for Eq.~(\ref{eq:GNLS}).
We previously benefited from merging the functionality of {\tt{SWtools}} and {\tt{py-fmas}} when studying the collision dynamics of localized solutions in a nonlinear optics GNSE \cite{Melchert:PRA:2024}. 
This allowed us to directly answer questions such as ``Is the nonlinear bound state a true soliton?'' (If they emerge from collisions with unchanged shapes and speeds, they can be considered solitons), 
and, ``What is the result of the interaction of solitary wave solutions in the considered model?''.
To allow users to proceed from idea to numerical experimentation to results as quickly as possible, we provide a complete workflow using {\tt{SWtools}} in conjunction with {\tt{py-fmas}} in Section~\ref{sec:FMAS}.

\paragraph{Extendibility and reusablity}
Finally, let us stress that the methods provided with {\tt{SWtools}} can  be extended to $d$-dimensional transverse coordinates $\vec{\xi}=(\xi_1, \ldots, \xi_d)$ and beyond the specific choice for $\hat{L}$ [see Eq.~(\ref{eq:L})] in straight forward manner: 
for the NSOM, the differential operators need to be expressed by suitable $d$-dimensional finite-difference approximants; in case of the SRM, the implemented spectral derivatives need to be extended to  the $d$-dimensions as well.
In this context, additional methods for the solution of the NEVP~(\ref{eq:EVP}) can be implemented by subclassing {\tt{SWtool}}'s base class {\tt{IterBase}}, allowing developers to reuse our tested software components.
%
%As supplementary material we provide an example, showing how to implement the explicit and fast converging ``accelerated imaginary time propagation method'' using the data structures provided by {\tt{SWtools}} \cite{suppMat}.
%
%As supplementary material we provide a brief manual containing a description of {\tt{IterBase}} and an example, showing how to implement the SRM for $d=2$ (see the example in Section~\ref{sec:SRM2D}).
% using the data structures provided by {\tt{SWtools}} \cite{suppMat}.
%
We refer user that aim to adapt {\tt{SWtools}} to his or her needs to the example in Section~\ref{sec:SRM2D}, demonstrating a custom $d=2$ SRM, and the online documentation~\cite{SWtools:GH:2025}.

\section{Conclusions\label{sec:conclusion}}

With the {\tt{SWtools}} Python package we present an extendible resource for researchers, i.e.\ developers and users alike, that face the scientific problem of solving NEVPs of the type of Eq.~(\ref{eq:EVP}).
%
%As solitons assume a leading role in the dynamics of general solutions of propagation equations, as, e.g., the GNSE~(\ref{eq:GNLS}), the ability to determine numerical solutions in absence of analytical solutions is of fundamental importance.
%
In closing, let us again emphasize that solitons assume a leading role in the dynamics of general solutions of propagation equations, as, e.g., the GNSE~(\ref{eq:GNLS}).
Since in most applicable cases exact soliton expressions are absent, the ability to determine such solutions numerically is of fundamental importance.
Soliton solutions calculated via {\tt{SWtools}} can then be analyzed further or used as initial conditions in pulse propagation simulations \cite{Melchert:CPC:2022,Melchert:SFX:2022}.
While our interest in the topic lies in the field of nonlinear optics, where optical solitons and soliton related phenomena do not cease to offer new and exciting perspectives~\cite{Redondo:NP:2023}, we hope to also spark the interest of researchers in other fields of science.
We further hope to benefit researchers and students that are looking for examples that guide their own numerical experimentation. 
% a tutorial-type introduction to the considered problems.
%
In this regard, the online documentation \cite{SWtools:GH:2025} and the minimal examples given in Sections~\ref{sec:1DBEC}-\ref{sec:SRM2D} can serve as classroom code or as starting point for seminar projects in computational courses. 

Possible directions for improving the presented software by extending its range of applicability include the implementation of further methods \cite{Lakoba:JCP:2007,Yang:SIAM:2007,Yang:SIAM:2008,Yang:JCP:2009,Lakoba:PD:2009,Lehtovaara:JCP:2007,Bao:SIAM:2004}, addressing, e.g., systems of coupled equations of the type of Eq.~(\ref{eq:GNLS}).
{\tt{SWtools}} is available for download under Ref.~\cite{SWtools:GH:2025}.

\section*{Acknowledgements}

We acknowledge support from the Deutsche Forschungsgemeinschaft  (DFG) under
Germany’s Excellence Strategy within the Cluster of Excellence PhoenixD
(Photonics, Optics, and Engineering – Innovation Across Disciplines) (EXC 2122,
projectID 390833453).

\appendix

\section{Documentation of subclass {\tt{SRM}}\label{sec:SRM_subclass}}
%\paragraph{Implementation details}
%
{\tt SWtools} provides the class {\tt{SRM}} that implements the method of  Section~\ref{sec:alg_02}, see Fig.~\ref{fig:pictorialOverview}(b).
Instantiating an instance of the class {\tt{SRM}} requires a user to specify several input parameters.
Below they are listed as ``{\tt{parameter\_name}} (data type): description'':
\begin{itemize}
\setlength{\itemsep}{0em}
\item {\tt{xi}} (array): Discretized coordinate $\xi$;
\item {\tt{cL}} (array): Coefficients $c_L=(c_1,c_2, c_3, c_4)$ defining the linear operator $\hat{L}$ [Eq.~(\ref{eq:L})];
\item {\tt{F}} (function): Nonlinear functional. Call signature  {\tt{F(I, xi)}} with intensity $I=|U|^2$ ({\tt{I}}), and coordinate $\xi$ ({\tt{xi}}). 
\item {\tt{tol}} (float): Iteration is stopped when the accuracy falls below this tolerance threshold (default: $10^{-12}$);
\item {\tt{maxiter}} (int): Maximum number of allowed iterations (default: $10^4$);
\item {\tt{nskip}} (int): Number of iterations to skip in between storing intermediate results (default: $1$);
\item {\tt{verbose}} (bool): Set to True to print details during iteration (default: False).
\end{itemize}
The class {\tt{SRM}} provides the following methods, listed in the format ``{\tt{method\_name}}(par1, par2, ...): description'':
\begin{description}
\item[{\tt{solve(U0, kap, **kwargs)}}]: Performs iterative solution of the NEVP~(\ref{eq:EVP_SRM}) via the SRM. Inherited from superclass {\tt{IterBase}}, see Fig.~\ref{fig:pictorialOverview}(a).
 \begin{description}
 \setlength{\itemsep}{0em}
    \item {\bf{Parameters}}:
    \begin{itemize}
    \setlength{\itemsep}{0em}
    \item {\tt{U0}} (array): Initial condition in Eq.~(\ref{eq:EVP_aux_ic});
    \item {\tt{kap}} (float): Eigenvalue $\kappa$;
    \end{itemize}
    \item {\bf{Optional keyword arguments}}:
    \begin{itemize}
    \setlength{\itemsep}{0em}
    \item {\tt{ortho\_set}}: List of previously found orthogonal solutions (default: empty list);
    \item {\tt{acc\_fun}}: Function for calculating the accuracy $\epsilon_n$ at step $n$. Call signature {\tt{acc\_fun(xi, Up, Uc)}} with coordinate $\xi$ ({\tt{xi}}), solution $U^{(n-1)}$ at the previous step ({\tt{Up}}), and solution $U^{(n)}$ at the current step ({\tt{Uc}}) (default: see Eq.~(\ref{eq:acc})).
    \end{itemize}
    \item {\bf{Returns}}:
    \begin{itemize}
    \setlength{\itemsep}{0em}
    \item {\tt{U}} (array): SRM solution $U^\star$;
    \item {\tt{acc}} (float): Terminal accuracy $\epsilon^\star$;
    \item {\tt{succ}} (bool): Boolean flag indicating if the iteration procedure exited successfully;
    \item {\tt{msg}} (str): Cause of the termination. 
    \end{itemize}
 \end{description}
\item[{\tt{functional\_N(U)}}]: Implements Eq.~(\ref{eq:IOM_N}).
\begin{description}
    \setlength{\itemsep}{0em}
    \item {\bf{Parameters}}:
    \begin{itemize}
    \item {\tt{U}} (array): Solution at the current iteration step.
    \end{itemize}
    \item {\bf{Returns}}:
        \begin{itemize}
    \item {\tt{N}} (float): Value of the functional $N[U]$.
    \end{itemize}
 \end{description}
\item[{\tt{functional\_H(U)}}]: Implements Eq.~(\ref{eq:IOM_H}) using spectral derivatives to handle $\hat{L}$.
\begin{description}
    \setlength{\itemsep}{0em}
    \item {\bf{Parameters}}:
    \begin{itemize}
    \item {\tt{U}} (array): Solution at the current iteration step.
    \end{itemize}
    \item {\bf{Returns}}:
        \begin{itemize}
    \item {\tt{H}} (float): Value of the functional $H[U]$.
    \end{itemize}
 \end{description}
\item[{\tt{singleUpdate(U, N, H, kap, **kwargs}}]: Implements a single step of the $d=1$ SRM detailed in Refs.~\cite{Ablowitz:OL:2005,Fibich:PD:2006}.
\begin{description}
    \setlength{\itemsep}{0em}
    \item {\bf{Parameters}}:
    \begin{itemize}
    \item {\tt{U}} (array): Solution at the current iteration step;
    \item {\tt{N}} (float): Current value of $N[U]$;
    \item {\tt{H}} (float): Current value of $H[U]$;
    \item {\tt{kap}} (float): Eigenvalue $\kappa$.
    \end{itemize}
    \item {\bf{Returns}}:
        \begin{itemize}
    \item {\tt{U}} (array): Updated solution.
    \end{itemize}
 \end{description}
\item[{\tt{show(f\_name='none')}}]: Produces a figure of the solution $U^\star$ and the convergence of the accuracy. 
%Designed for models where $\xi$ is one-dimensional.
\begin{description}
    \setlength{\itemsep}{0em}
    \item {\bf{Parameters}}:
    \begin{itemize}
    \item {\tt{f\_name}} (str): Figure name. If {\tt{f\_name}} is not set, the figure is displayed directly.
    \end{itemize}
    \item {\bf{Returns}}: None
 \end{description}
 \end{description}

\section{Documentation of subclass {\tt{NSOM}}\label{sec:NSOM_subclass}}
%\paragraph{Implementation details}
%
{\tt SWtools} provides a class {\tt{NSOM}} that implements the method of Section~\ref{sec:alg_01}, see Fig.~\ref{fig:pictorialOverview}(c).
Instantiating an instance of the class {\tt{NSOM}} requires a user to specify several input parameters.
Below they are listed as ``{\tt{parameter\_name}} (data type): description'':
\begin{itemize}
\setlength{\itemsep}{0em}
\item {\tt{xi}} (array): Discretized coordinate $\xi$;
\item {\tt{cL}} (array): Coefficients $c_L=(c_1,c_2, c_3, c_4)$ defining the linear operator $\hat{L}$;
\item {\tt{F}} (function): Nonlinear functional. Call signature  {\tt{F(I, xi)}} with intensity $I=|U|^2$ ({\tt{I}}), and coordinate $\xi$ ({\tt{xi}}). 
\item {\tt{ORP}} (float): Overrelaxation parameter (default: $1$);
\item {\tt{tol}} (float): Iteration is stopped when the accuracy falls below this tolerance threshold (default: $10^{-12}$);
\item {\tt{maxiter}} (int): Maximum number of allowed iterations (default: $10^4$);
\item {\tt{nskip}} (int): Number of iterations to skip in between kept intermediate results (default: $1$);
\item {\tt{verbose}} (bool): Set to True to print details during iteration (default: False).
\end{itemize}
The class {\tt{NSOM}} provides the following methods, listed in the format ``{\tt{method\_name}}(par1, par2, ...): description'':
\begin{description}
\item[{\tt{solve(U0, N0, **kwargs)}}]: Performs iterative solution of the NEVP~(\ref{eq:EVP}), supplemented by the normalization constraint $N[U]=N_0$ [Eq.~(\ref{eq:EVP_aux_N})]. Inherited from superclass {\tt{IterBase}}, see Fig.~\ref{fig:pictorialOverview}(a).
 \begin{description}
 \setlength{\itemsep}{0em}
    \item {\bf{Parameters}}:
    \begin{itemize}
    \setlength{\itemsep}{0em}
    \item {\tt{U0}} (array): Initial condition in Eq.~(\ref{eq:EVP_aux_ic});
    \item {\tt{N0}} (float): Value for constraint~(\ref{eq:EVP_aux_N});
    \end{itemize}
    \item {\bf{Optional keyword arguments}}:
    \begin{itemize}
    \setlength{\itemsep}{0em}
    \item {\tt{ortho\_set}}: List of previously found orthogonal solutions (default: empty list);
    \item {\tt{acc\_fun}}: Function for calculating the accuracy $\epsilon_n$ at step $n$. Call signature {\tt{acc\_fun(xi, Up, Uc)}} with coordinate $\xi$ ({\tt{xi}}), solution $U^{(n-1)}$ at the previous step ({\tt{Up}}), and solution $U^{(n)}$ at the current step ({\tt{Uc}}) (default: see Eq.~(\ref{eq:acc})).
    \end{itemize}
    \item {\bf{Returns}}:
    \begin{itemize}
    \setlength{\itemsep}{0em}
    \item {\tt{U}} (array): SRM solution $U^\star$;
    \item {\tt{acc}} (float): Terminal accuracy $\epsilon^\star$;
    \item {\tt{succ}} (bool): Boolean flag indicating if the iteration procedure exited successfully;
    \item {\tt{msg}} (str): Cause of the termination. 
    \end{itemize}
 \end{description}
\item[{\tt{functional\_N(U)}}]: Implements Eq.~(\ref{eq:IOM_N}).
\begin{description}
    \setlength{\itemsep}{0em}
    \item {\bf{Parameters}}:
    \begin{itemize}
    \item {\tt{U}} (array): Solution at the current iteration step.
    \end{itemize}
    \item {\bf{Returns}}:
        \begin{itemize}
    \item {\tt{N}} (float): Value of the functional $N[U]$.
    \end{itemize}
 \end{description}
\item[{\tt{functional\_H(U)}}]: Implements Eq.~(\ref{eq:IOM_H}) using a five-point finite-difference approximation of $\hat{L}$.
\begin{description}
    \setlength{\itemsep}{0em}
    \item {\bf{Parameters}}:
    \begin{itemize}
    \item {\tt{U}} (array): Solution at the current iteration step.
    \end{itemize}
    \item {\bf{Returns}}:
        \begin{itemize}
    \item {\tt{H}} (float): Value of the functional $H[U]$.
    \end{itemize}
 \end{description}
\item[{\tt{singleUpdate(U, N, H, N0, **kwargs)}}]: Implements a single step of a nonlinear successive overrelaxation method (NSOM) \cite{Langtangen:BOOK:2019} for the NEVP~(\ref{eq:EVP}). The procedure uses a Gauss-Seidel method with overrelaxation to iteratively updata a user-provided initial condition in place \cite{Press:BOOK:2007,Langtangen:BOOK:2017}.
\begin{description}
    \setlength{\itemsep}{0em}
    \item {\bf{Parameters}}:
    \begin{itemize}
    \item {\tt{U}} (array): Solution at the current iteration step;
    \item {\tt{N}} (float): Current value of $N[U]$;
    \item {\tt{H}} (float): Current value of $H[U]$;
    \item {\tt{N0}} (float): Normalization constraint $N0$.
    \end{itemize}
    \item {\bf{Returns}}:
        \begin{itemize}
    \item {\tt{U}} (array): Updated solution.
    \end{itemize}
 \end{description}
\item[{\tt{show(f\_name='none')}}]: Produces a figure of the solution $U^\star$ and the convergence of the accuracy. 
%Designed for models where $\xi$ is one-dimensional.
\begin{description}
    \setlength{\itemsep}{0em}
    \item {\bf{Parameters}}:
    \begin{itemize}
    \item {\tt{f\_name}} (str): Figure name. If {\tt{f\_name}} is not set, the figure is displayed directly.
    \end{itemize}
    \item {\bf{Returns}}: None
 \end{description}
 \end{description}

%Numerical experiments that verify the implementation are reported in Sect.~(\ref{sec:verification}).

%\clearpage

%% The Appendices part is started with the command \appendix;
%% appendix sections are then done as normal sections
%% \appendix

%% \section{}
%% \label{}

%% References
%%
%% Following citation commands can be used in the body text:
%% Usage of \cite is as follows:
%%   \cite{key}         ==>>  [#]
%%   \cite[chap. 2]{key} ==>> [#, chap. 2]
%%

%% References with bibTeX database:

\bibliographystyle{elsarticle-num}
\bibliography{references}

\begin{thebibliography}{10}
\expandafter\ifx\csname url\endcsname\relax
  \def\url#1{\texttt{#1}}\fi
\expandafter\ifx\csname urlprefix\endcsname\relax\def\urlprefix{URL }\fi
\expandafter\ifx\csname href\endcsname\relax
  \def\href#1#2{#2} \def\path#1{#1}\fi

\bibitem{Zabusky:PRL:1965}
N.~J. Zabusky, M.~D. Kruskal, Interaction of "solitons" in a collisionless
  plasma and the recurrence of initial states, Phys. Rev. Lett. 15 (1965)
  240--243.

\bibitem{Zakharov:JETP:1972}
V.~E. Zakharov, A.~B. Shabat, {Exact theory of two-dimensional self-focusing
  and one-dimensional self-modulation of waves in nonlinear media}, Sov. Phys.
  JETP 34 (1972) 62.

\bibitem{Drazin:BOOK:1989}
P.~G. Drazin, R.~S. Johnson, Solitons: An Introduction, Cambridge University
  Press, 1989.

\bibitem{Kivshar:RMP:1989}
Y.~S. Kivshar, {Dynamics of solitons in nearly integrable systems}, Rev. Mod.
  Phys. 61 (1989) 763.

\bibitem{Frauenkron:PRE:1996}
H.~Frauenkron, Y.~S. Kivshar, B.~A. Malomed, Multisoliton collisions in nearly
  integrable systems, Phys. Rev. E 54 (1996) R2244.

\bibitem{Jakubowski:PRE:1997}
M.~H. Jakubowski, K.~Steiglitz, R.~Squier, {Information transfer between
  solitary waves in the saturable Schr\"odinger equation}, Phys. Rev. E 56
  (1997) 7267.

\bibitem{Anastassiou:PRL:1999}
C.~Anastassiou, M.~Segev, K.~Steiglitz, J.~A. Giordmaine, M.~Mitchell, M.-F.
  Shih, S.~Lan, J.~Martin, Energy-exchange interactions between colliding
  vector solitons, Phys. Rev. Lett. 83 (1999) 2332.

\bibitem{Yang:PRL:2000}
J.~Yang, Y.~Tan, {Fractal Structure in the Collision of Vector Solitons}, Phys.
  Rev. Lett. 85 (2000) 3624.

\bibitem{Dmitriev:Chaos:2002}
S.~V. Dmitriev, T.~Shigenari, {Short-lived two-soliton bound states in weakly
  perturbed nonlinear Schrödinger equation}, Chaos 12 (2002) 324.

\bibitem{Feigenbaum:OE:2004}
E.~Feigenbaum, M.~Orenstein, Colored solitons interactions: particle-like and
  beyond, Opt. Express 12 (2004) 2193.

\bibitem{Goodman:PRL:2007}
R.~H. Goodman, R.~Haberman, {Chaotic Scattering and the $n$-Bounce Resonance in
  Solitary-Wave Interactions}, Phys. Rev. Lett. 98 (2007) 104103.

\bibitem{Edmonds:NJP:2017}
M.~J. Edmonds, T.~Bland, R.~Doran, N.~G. Parker, {Engineering bright
  matter-wave solitons of dipolar condensates}, New Journal of Physics 19
  (2017) 023019.

\bibitem{Dingwall:NJP:2018}
R.~J. Dingwall, M.~J. Edmonds, J.~L. Helm, B.~A. Malomed, P.~Öhberg,
  Non-integrable dynamics of matter-wave solitons in a density-dependent gauge
  theory, New Journal of Physics 20 (2018) 043004.

\bibitem{Melchert:PRL:2019}
O.~Melchert, S.~Willms, S.~Bose, A.~Yulin, B.~Roth, F.~Mitschke, U.~Morgner,
  I.~Babushkin, A.~Demircan, Soliton molecules with two frequencies, Phys. Rev.
  Lett. 123 (2019) 243905.

\bibitem{Rao:PRE:2020}
J.~Rao, J.~He, T.~Kanna, D.~Mihalache, {Nonlocal $M$-component nonlinear
  Schr\"odinger equations: Bright solitons, energy-sharing collisions, and
  positons}, Phys. Rev. E 102 (2020) 032201.

\bibitem{Melchert:PRA:2024}
O.~Melchert, A.~Demircan, Numerical investigation of solitary-wave solutions
  for the nonlinear schrödinger equation perturbed by third-order and negative
  fourth-order dispersion, Phys. Rev. A 110 (2024) 043518.

\bibitem{Paredes:PD:2020}
A.~Paredes, D.~N. Olivieri, HumbertoMichinel, {From optics to dark matter: A
  review on nonlinear Schrödinger–Poisson systems}, Physica D: Nonlinear
  Phenomena 403 (2020) 132301.

\bibitem{Zagorac:PRD:2022}
J.~L. Zagorac, I.~Sands, N.~Padmanabhan, R.~Easther, Schr\"odinger-poisson
  solitons: Perturbation theory, Phys. Rev. D 105 (2022) 103506.

\bibitem{Dalfovo:RMP:1999}
F.~Dalfovo, S.~Giorgini, L.~P. Pitaevskii, S.~Stringari, {Theory of
  Bose-Einstein condensation in trapped gases}, Rev. Mod. Phys. 71 (1999)
  463--512.

\bibitem{Mitschke:BOOK:2016}
F.~Mitschke, Fiber Optics: Physics and Technology, Springer, 2016.

\bibitem{Redondo:NP:2023}
A.~Blanco-Redondo, C.~M. de~Sterke, C.~Xu, S.~Wabnitz, S.~K. Turitsyn, {The
  bright prospects of optical solitons after 50 years}, Nat. Photon. 17 (2023)
  937.

\bibitem{Gardner:PRL:1967}
C.~Gardner, J.~Greene, M.~Kruskal, R.~Miura, {Method for solving the
  Korteweg-de Vries equation}, Phys. Rev. Lett. 19 (1967) 1095.

\bibitem{Lax:CPAM:1968}
P.~Lax, {Integrals of nonlinear equations of evolution and solitary waves},
  Comm. Pure Appl. Math. 21 (1968) 467.

\bibitem{Ablowitz:PRL:1973}
M.~J. Ablowitz, D.~J. Kaup, A.~C. Newell, H.~Segur, {Nonlinear-Evolution
  Equations of Physical Significance}, Phys. Rev. Lett. 31 (1973) 125.

\bibitem{Ablowitz:SAM:1974}
M.~Ablowitz, D.~Kaup, A.~Newell, H.~Segur, {The inverse scattering transform -
  Fourier analysis for nonlinear problems}, Studies Appl. Math. 53 (1974) 249.

\bibitem{Satsuma:PTP:1974}
J.~Satsuma, N.~Yajima, {Initial Value Problems of One-Dimensional
  Self-Modulation of Nonlinear Waves in Dispersive Media}, Prog. Theor. Phys.
  (Suppl.) 55 (1974) 284.

\bibitem{Kaup:JMP:1975}
D.~J. Kaup, {Exact quantization of the nonlinear Schrödinger equation}, J.
  Math. Phys. 16 (1975) 2036.

\bibitem{Miles:SIAM:1981}
J.~W. Miles, {An envelope soliton problem}, SIAM 41 (1981) 227--230.

\bibitem{Press:BOOK:2007}
W.~Press, S.~Teukolsky, W.~Vetterling, B.~Flannery, {Numerical Recipes: The Art
  of Scientific Computing}, Cambridge University Press, 2007.

\bibitem{Langtangen:BOOK:2019}
H.~P. Langtangen, K.-A. Mardal, {Introduction to Numerical Methods for
  Variational Problems}, Springer, 2019.

\bibitem{Schechter:AMS:1962}
S.~Schechter, {Iteration Methods for Nonlinear Problems}, Trans. Amer. Math.
  Soc. 104 (1962) 179--189.

\bibitem{Ortega:SIAM:1966}
J.~M. Ortega, M.~L. Rockoff, {Nonlinear Difference Equations and Gauss-Seidel
  Type Iterative Methods}, J. SIAM Numer. Anal. 3 (1966) 497--513.

\bibitem{Brewster:NM:1984}
M.~E. Brewster, R.~Kannan, {Nonlinear Successive Over-Relaxation}, Numer. Math.
  44 (1984) 309--315.

\bibitem{Yang:BOOK:2010}
J.~Yang, {Nonlinear Waves in Integrable and Nonintegrable Systems}, SIAM, 2010.

\bibitem{Haelterman:PRE:1994}
M.~Haelterman, A.~Sheppard, Bifurcation phenomena and multiple soliton-bound
  states in isotropic kerr media, Phys. Rev. E 149 (1994) 3376.

\bibitem{Mitchell:PRL:1997}
M.~Mitchell, M.~Segev, T.~Coskun, D.~Christodulides, Theory of self-trapped
  spatially incoherent light beams, Phys. Rev. Lett. 79 (1997) 4990.

\bibitem{Yang:PRE:2002}
J.~Yang, Z.~Chen, Defect solitons in photonic lattices, Phys. Rev. E 73 (2006)
  026609.

\bibitem{Petviashvili:SJPP:1976}
V.~I. Petviashvili, Equation for an extraordinary soliton, Sov. J. Plasma Phys.
  2 (1976) 257.

\bibitem{Lakoba:JCP:2007}
T.~Lakoba, J.~Yang, A generalized petviashvili iteration method for scalar and
  vector hamiltonian equations with arbitrary form of nonlinearity, J. Comp.
  Phys. 226 (2007) 1668.

\bibitem{Bao:SIAM:2004}
W.~Bao, Q.~Du, {Computing the ground state solution of Bose-Einstein
  condensates by a normalized gradient flow}, SIAM J. Sci. Comput. 25 (2004)
  1674--1697.

\bibitem{Lehtovaara:JCP:2007}
L.~Lehtovaara, J.~Toivanen, J.~Eloranta, {Solution of the time-independent
  Schrödinger equation by the imaginary time propagation method}, J. Comp.
  Phys. 221 (2007) 148--157.

\bibitem{Yang:SIAM:2008}
J.~Yang, T.~I. Lakoba, {Accelerated Imaginary-time Evolution Methods for the
  Computation of Solitary Waves}, Studies in Applied Mathematics 120 (2008)
  265--292.

\bibitem{Yang:SIAM:2007}
J.~Yang, T.~I. Lakoba, {Universally-Convergent Squared-Operator Iteration
  Methods for Solitary Waves in General Nonlinear Wave Equations}, Studies in
  Applied Mathematics 118 (2007) 153--197.

\bibitem{Boyd:JCP:2002}
J.~P. Boyd, {Deleted Residuals, the QR-Factored Newton Iteration, and Other
  Methods for Formally Overdetermined Determinate Discretizations of Nonlinear
  Eigenproblems for Solitary, Cnoidal, and Shock Waves}, Journal of
  Computational Physics 179 (2002) 216--237.

\bibitem{Maytee:PRA:2006}
T.~Mayteevarunyoo, B.~A. Malomed, {Stability limits for gap solitons in a
  Bose-Einstein condensate trapped in a time-modulated optical lattice}, Phys.
  Rev. A 74 (2006) 033616.

\bibitem{Melchert:SFX:2021}
O.~Melchert, A.~Demircan, {pyGLLE: A Python toolkit for solving the generalized
  Lugiato–Lefever equation}, SoftwareX 15 (2021) 100741.

\bibitem{Yang:JCP:2009}
J.~Yang, Newton-conjugate-gradient methods for solitary wave computations, J.
  Comp. Phys. 228 (2009) 7007.

\bibitem{Lakoba:PD:2009}
T.~Lakoba, Conjugate gradient method for finding fundamental solitary waves,
  Physica D 238 (2009) 2308.

\bibitem{Rose:PD:1988}
H.~A. Rose, M.~I. Weinstein, {On the bound-states of the nonlinear Schrödinger
  equation with a linear potential}, Physica D 30 (1988) 207--218.

\bibitem{Zakharov:JETP:1998}
V.~E. Zakharov, E.~A. Kuznetsov, {Optical solitons and quasisolitons}, JETP 86
  (1998) 1035--1046.

\bibitem{Tsoy:PRA:2024}
E.~N. Tsoy, L.~A. Suyunov, {Generic quartic solitons in optical media}, Phys.
  Rev. A 109 (2024) 053528.

\bibitem{Besse:BOOK:2015}
C.~Besse, J.-C. Garreau, {Nonlinear Optical and Atomic Systems}, Springer,
  2015.

\bibitem{Agrawal:BOOK:2019}
G.~P. Agrawal, Nonlinear Fiber Optics, Academic Press, 2019.

\bibitem{Ablowitz:PD:2003}
M.~J. Ablowitz, Z.~H. Musslimani, {Discrete spatial solitons in a
  diffraction-managed nonlinear waveguide array: a unified approach}, Physica D
  184 (2003) 276–303.

\bibitem{Ablowitz:OL:2005}
M.~J. Ablowitz, Z.~H. Musslimani, Spectral renormalization method for computing
  self-localized solutions to nonlinear systems, Opt. Lett. 30 (2005) 2140.

\bibitem{Musslimani:JOSAB:2004}
Z.~Musslimani, J.~Yang, Self-trapping of light in a two-dimensional photonic
  lattice, J. Opt. Soc. Am. B 21 (2004) 973.

\bibitem{Ablowitz:EPJST:2009}
M.~Ablowitz, T.~Horikis, Solitons and spectral renormalization methods in
  nonlinear optics, Eur. Phys. J. Spec. Top. 173 (2009) 147.

\bibitem{Bao:JCP:2003}
W.~Bao, W.~Tang, {Ground-state solution of Bose–Einstein condensate by
  directly minimizing the energy functional}, Journal of Computational Physics
  187 (2003) 230--254.

\bibitem{Liu:SIAM:2021}
W.~Liu, Y.~Cai, {Normalized Gradient Flow with Lagrange Multiplier for
  Computing Ground States of Bose–Einstein Condensates}, SIAM J. Sci. Comput.
  43 (2021) B219--B242.

\bibitem{SWtools:GH:2025}
O.~Melchert, {{\tt SWtools} -- Iterative solvers for solitary wave solutions of
  nonlinear Schrödinger-type equations},
  \url{https://github.com/omelchert/SWtools.git}, [Online; accessed 2025-04-09]
  (2025).

\bibitem{Rossum:1995}
G.~Rossum, {Python Reference Manual}, Tech. rep., Amsterdam, The Netherlands,
  The Netherlands (1995).

\bibitem{Virtanen:NM:2020}
P.~Virtanen, R.~Gommers, T.~E. e.~a. Oliphant, {SciPy 1.0: fundamental
  algorithms for scientific computing in Python}, Nature Methods 17 (2020) 261.

\bibitem{Hunter:CSE:2007}
J.~D. Hunter, {Matplotlib: A 2D graphics environment}, Computing in Science \&
  Engineering 9 (2007) 90.

\bibitem{Fibich:PD:2006}
G.~Fibich, Y.~Sivan, M.~I. Weinstein, {Bound states of nonlinear Schrödinger
  equations with a periodic nonlinear microstructure}, Physica D 217 (2006) 31.

\bibitem{Langtangen:BOOK:2017}
H.~P. Langtangen, S.~Linge, {Finite-Difference Computing with PDEs}, Springer,
  2017.

\bibitem{Melchert:CPC:2022}
O.~Melchert, A.~Demircan, {py-fmas: A python package for ultrashort optical
  pulse propagation in terms of forward models for the analytic signal},
  Computer Physics Communications 273 (2022) 108257.

\bibitem{Karlsson:OC:1994}
M.~Karlsson, A.~H\"o\"ok, {Soliton-like pulses governed by fourth order
  dispersion in optical fibers}, Optics Communications 104~(4) (1994) 303--307.

\bibitem{Akhmediev:OC:1994}
N.~Akhmediev, A.~Buryak, M.~Karlsson, Radiationless optical solitons with
  oscillating tails, Opt. Commun. 110 (1994) 540.

\bibitem{Piche:OL:1996}
M.~Pich\'e, J.-F. Cormier, X.~Zhu, {Bright optical soliton in the presence of
  fourth-order dispersion}, Opt. Lett. 21 (1996) 845.

\bibitem{deSterke:OC:2023}
C.~M. {de Sterke}, A.~Blanco-Redondo, {Even-order dispersion solitons: A
  pedagogical note}, Optics Communications 541 (2023) 129560.

\bibitem{Sinkin:JLT:2003}
O.~Sinkin, R.~Holzlohner, J.~Zweck, C.~Menyuk, {Optimization of the split-step
  Fourier method in modeling optical-fiber communications systems}, Journal of
  Lightwave Technology 21 (2003) 61--68.

\bibitem{Yang:OL:2003}
J.~Yang, Z.~H. Musslimani, Fundamental and vortex solitons in a two-dimensional
  optical lattice, Opt. Lett. 28~(21) (2003) 2094--2096.

\bibitem{Ostrovskaya:OE:2004}
E.~A. Ostrovskaya, Y.~S. Kivshar, {Photonic crystals for matter waves:
  Bose-Einstein condensates in optical lattices}, Opt. Express 12~(1) (2004)
  19--29.

\bibitem{Melchert:arxiv:2024}
O.~Melchert, A.~Demircan, {Optical Solitary Wavelets}, see
  \href{https://doi.org/10.48550/arXiv.2410.06867}{arXiv.2410.06867} (2024).

\bibitem{Melchert:OPTIK:2023}
O.~Melchert, S.~Willms, I.~Babushkin, U.~Morgner, A.~Demircan, {(Invited)
  Two-color soliton meta-atoms and molecules}, Optik 280 (2023) 170772.

\bibitem{Tam:PRA:2020}
K.~K.~K. Tam, T.~J. Alexander, A.~Blanco-Redondo, C.~M. de~Sterke, {Generalized
  dispersion Kerr solitons}, Phys. Rev. A 101 (2020) 043822.

\bibitem{Melchert:OL:2023}
O.~Melchert, S.~Bose, S.~Willms, I.~Babushkin, U.~Morgner, A.~Demircan,
  Two-color pulse compounds in waveguides with a zero-nonlinearity point, Opt.
  Lett. 48 (2023) 518--521.

\bibitem{Weinstein:BOOK:2015}
C.~E. Wayne, M.~I. Weinstein, {Dynamics of Partial Differential Equations},
  Springer, 2015.

\bibitem{Melchert:SFX:2022}
O.~Melchert, A.~Demircan, {GNLStools.py: A generalized nonlinear Schrödinger
  Python module implementing different models of input pulse quantum noise},
  SoftwareX 20 (2022) 101232.

\end{thebibliography}

%% Authors are advised to submit their bibtex database files. They are
%% requested to list a bibtex style file in the manuscript if they do
%% not want to use elsarticle-num.bst.

%% References without bibTeX database:

% \begin{thebibliography}{00}

%% \bibitem must have the following form:
%%   \bibitem{key}...
%%

% \bibitem{}

% \end{thebibliography}

\end{document}